\documentclass{emulateapj}
\usepackage{graphicx}
\usepackage{amsmath}

\usepackage{color}

\newcommand{\appropto}{\mathrel{\vcenter{
  \offinterlineskip\halign{\hfil$##$\cr
    \propto\cr\noalign{\kern2pt}\sim\cr\noalign{\kern-2pt}}}}}

\def\kms{kms$^{-1}$\,}

\def\kms{\ifmmode{\rm km\thinspace s^{-1}}\else km\thinspace s$^{-1}$\fi}

\shortauthors{Rappaport et al.~2013}
\shorttitle{Thermally Bloated White Dwarfs}

\begin{document}

% ------------------------------------------------------------------------
% New commands
%
\def\ltsima{$\; \buildrel < \over \sim \;$}
\def\lsim{\lower.5ex\hbox{\ltsima}}
\def\gtsima{$\; \buildrel > \over \sim \;$}
\def\gsim{\lower.5ex\hbox{\gtsima}}
% -------------------------------------------------------------------------
%

\bibliographystyle{apj}

\title{
Discovery of Two New Thermally Bloated Low-Mass White Dwarfs Among the {\em Kepler} Binaries
}

\author{
S.~Rappaport\altaffilmark{1},
L.~Nelson\altaffilmark{2},
A.~Levine\altaffilmark{3},
R.~Sanchis-Ojeda\altaffilmark{1},
D.~Gandolfi\altaffilmark{4}, 
G.~Nowak\altaffilmark{5,6}
E.~Palle\altaffilmark{5,6}, and
A.~Prsa\altaffilmark{7}
}

\altaffiltext{1}{Physics Department and Kavli Institute for
  Astrophysics and Space Research, Massachusetts Institute of
  Technology, Cambridge, MA 02139, USA; sar@mit.edu; rsanchis86@gmail.com}
  \altaffiltext{2}{Department of Physics, Bishop's University, 2600 College
St., Sherbrooke, QC J1M 1Z7; lnelson@ubishops.ca}
\altaffiltext{3}{Kavli Institute for
  Astrophysics and Space Research, Massachusetts Institute of
  Technology, Cambridge, MA 02139, USA; aml@space.mit.edu}
\altaffiltext{4}{Landessternwarte K\"onigstuhl, Zentrum f\"ur Astronomie der Universit\"at Heidelberg, K\"onigstuhl 12, D-69117 Heidelberg, Germany; davide.gandolfi@lsw.uni-heidelberg.de}
\altaffiltext{5}{Instituto de Astrof\'isica de Canarias, C/ v\'ia L\'actea, s/n, E-38205 La Laguna, Tenerife, Spain; gnowak@iac.es}
\altaffiltext{6}{Departamento de Astrof\'isica, Universidad de La Laguna, Av.~Astrof\'isico Francisco S\'anchez, s/n, E-38206 La Laguna, Tenerife, Spain; epalle@iac.es}
\altaffiltext{7}{Villanova University, Villanova, PA 19085, USA; aprsa@villanova.edu}

\begin{abstract}

We report the discovery of two new low-mass, thermally bloated, hot white dwarfs among the 
{\em Kepler} sample of eclipsing binaries.  These are KIC 9164561 and KIC 10727668
with orbital periods of 1.2670 and 2.3058 days, respectively.  The current primary in 
both systems is an A star of about 2 $M_\odot$.  This brings the number of similar binaries 
among the {\em Kepler} sample to six, and the two new systems have the shortest orbital 
periods among them.  The white dwarf in KIC 9164561 has the largest thermal bloating, 
compared to its cold degenerate radius, of about a factor of 14.  We utilize RV measurements
of the A star in KIC 9164561 to determine the white dwarf mass rather accurately: $0.197 \pm 
0.005 \,M_\odot$.  The mass of the white dwarf in KIC 10727668 is based on the Doppler boosting signal
in the {\em Kepler} photometry, and is less accurately determined to be $0.266 \pm 0.035 \, M_\odot$. 
Based on the inferred radii and effective temperatures of these two white dwarfs we are able to make an {\em independent} theoretical estimate of their masses to within $\sim$0.01 $M_\odot$ based on evolutionary models of their cooling history after they lose their hydrogen-rich envelopes.  We also present evidence that there is a 
third body in the KIC 9164561 system with an orbital period of $8-14$ years.

\end{abstract}


\section{Introduction}

In relatively close primordial binaries which contain at least one star that can 
evolve substantially by the current epoch,  
mass transfer from the primary to the secondary is assured.  The
course of the evolution of each system depends on its initial orbital period, the two
stellar masses, whether the binary is part of a triple system, spin and tidal 
interactions, and so forth.  Some of the possible evolutionary channels have been explored and
even systematically mapped out by Rappaport et al.~(1994), Iben \& Tutukov (1995),
Nelson \& Eggleton (2001), Han et al.~(2002; 2003), Podsiadlowski et al.~(2003a), and Nelemans (2010).  
Depending largely upon the mass ratio and the initial orbital 
period, the mass transfer may be stable and mostly conservative, stable and mostly 
non-conservative, or dynamically unstable.  

In the case of a close binary that comprises a primordial primary star with mass $M_{1,0} \lesssim 2.5 \, 
M_\odot$ and a secondary star that has $M_{2,0} \gtrsim 0.7 \, M_\odot$, the mass
transfer is most likely to be stable.  However, depending on the ratio of thermal timescales of the two
stars, the mass transfer may range from highly non-conservative for small
$M_{2,0}/M_{1,0}$ ratios to mostly conservative for values of $M_{2,0}/M_{1,0}$ close to unity.  Whether the
orbit ultimately evolves to be longer or shorter than the original orbital period depends on
how much specific angular momentum is carried away with the matter ejected 
from the binary.  

In principle the best way to model such binary evolution is via a stellar evolution
code that can simultaneously handle the nuclear and mass loss/gain evolution of {\em both}
stars as well as the orbital evolution.  The recently developed {\it MESA} code (Paxton et al.~2011) 
is capable of doing such binary evolution studies, but, to our knowledge, no systematic studies have
yet been carried out of ordinary close primordial binaries.

We know empirically, and from some theoretical studies, that such evolution can lead
to systems such as Algol binaries (see, e.g., Nelson \& Eggleton 2001) where the 
mass transfer is stable and the envelope of the primary has not yet been exhausted; sdB 
binaries (see, e.g., Han et al.~2002; 2003) which may have evolved from
unstable mass transfer in a common envelope; systems with hot, bloated, low-mass
white dwarfs (see, e.g., van Kerkwijk et al.~2010; Carter et al.~2011), where the envelope of the primary has been completely removed; and even double-degenerate binaries (Marsh et al.~2004; Iben, Tutukov, \& Yungelson 1995; Nelemans 2010).

In low-mass binaries where the mass transfer proceeds stably, regardless of whether it is or is not conservative,
there is a fairly tight relation between the white dwarf mass, $M_{\rm wd}$, and the 
final orbital period, $P_{\rm orb}$ (see, e.g., Joss \& Rappaport 1983; Savonije 1983; Paczy\'nski 1983;
Savonije 1987; Rappaport et al.~1995; Fig.~5 of Lin et al.~2011; Tauris \& van den Heuvel 2014).
For orbital periods $2~{\rm days} \lesssim P_{\rm orb} \lesssim$ years, and 
white dwarf masses, $M_{\rm wd}$, up to 1.4 $M_\odot$, in systems where the initial masses of the primaries, $M_{1,0}$, were no larger than
$\sim$$2.5 \, M_\odot$,
the following expression gives the final orbital period in terms of the white dwarf mass:
\begin{equation}
P_{\rm orb} \simeq \frac{1.3 \times 10^5 M_{\rm wd}^{6.25}}{\left(1+4 M_{\rm wd}^4\right)^{1.5}} ~{\rm days~~.}
\label{eqn:Mwd_P}
\end{equation}
For orbital periods up to $\sim$20 days, which apply to systems with  
primaries of initial masses up to $\sim$$2.5 \, M_\odot$, a more recently derived
expression (Lin et al.~2011) is:
\begin{equation}
P_{\rm orb} \simeq \frac{4.6 \times 10^6 ~M_{\rm wd}^9}{(1+25 M_{\rm wd}^{3.5} + 29M_{\rm wd}^6)^{3/2}}{\rm ~days} ~~.
\label{eqn:Mwd_P2}
\end{equation}
These expressions have been checked against a number of binary radio pulsars (Fig.~3 of Rappaport 
et al.~1995; Fig.~2 of Tauris \& van den Heuvel 2014), and against a number of {\em Kepler}
binaries containing low-mass hot white dwarfs (van Kerkwijk et al.~2010; Fig.~6 of Carter et al.~2011).  

Prior to this work, 
only four low-mass white dwarf systems in the {\em Kepler} field that can 
be used to study this relation between orbital period and white dwarf mass were identified.  They are KOI 74,
KOI 81 (Rowe et al.~2010; van Kerkwijk et al.~2010), KHWD3 (KOI 1375; Carter et al.~2011), 
and KOI 1224 (Breton et al.~2012).  In this work we report the discovery of 
new low-mass, highly thermally bloated white dwarfs in the {\em Kepler} binaries 
KIC 9164561 and KIC 10727668.  The orbital periods of these binaries are 
1.267 and 2.305 days, the shortest periods among the six {\it Kepler} binaries containing low-mass 
hot white dwarfs.  The white dwarf in KIC 9164561 is by far the most thermally bloated of any of the 
six white dwarfs.

In Section \ref{sec:data} we discuss the search that led to these two systems among
the {\em Kepler} binaries, and indicate how the data were prepared for analysis.  Folded
light curves are presented in Section \ref{sec:LCs} along with a simple geometric analysis, 
including a harmonic decomposition of the out-of-eclipse portions of the light curve.  The masses of 
the primary A stars are inferred from information on their temperatures and mean densities in Sect.~\ref{sec:Astars}.
In Sect.~\ref{sec:RV} we present the radial velocity and spectral studies of KIC 9164561.  The mass ratio of KIC 10727668 is examined in the context of the Doppler boosting effect in Sect.~\ref{sec:DB}.  Inferences about the mass
ratios in the two systems are drawn from evaluating the ellipsoidal light variations in Sect.~\ref{sec:ELV}.
In Sect.~\ref{sec:WDs} the system parameters for these two binaries are evaluated with emphasis 
on the white dwarf properties.  Evidence for a third star in KIC 9164561 is presented in Sect.~\ref{sec:third}. Finally, in Sect.~\ref{sec:discuss} we discuss KIC 9164561 and KIC 10727668 in the context of different binary evolution scenarios and white dwarf cooling models.

\section{Data Preparation and Search}
\label{sec:data}

In the process of searching for third-body companions to the binary stars in the {\em Kepler}
sample (see, e.g., Rappaport et al.~2013; see also Conroy et al.~2014), we produced both an $O-C$ curve for 
the eclipses (``observed minus computed'' eclipse times) as well as a folded light
curve for each of the $\sim$2600 cataloged {\em Kepler} binaries (Matijevic et al.~2012; Kirk et al.~2015)\footnote{
\url{http://keplerebs.villanova.edu/}}.  We
used the {\em Kepler} long-cadence time-series photometric data (29.4~min samples) 
obtained from quarters 0 through 17. For all the known binary systems for which at least 
one quarter of photometry is available, the version 5.0 FITS files, which are available for 
all quarters, were downloaded from the STScI MAST website\footnote{\url{https://archive.stsci.edu/kepler/}}. 
We used the data that had 
been processed with the PDC-MAP algorithm (Stumpe et al.~2012; Smith et al.~2012), which 
is designed to remove many instrumental artifacts from the time series while preserving any 
astrophysical variability. We also made use of the information in the file headers to obtain 
estimates of the combined differential photometric precision (CDPP).

We then examined by eye all 2600 $O-C$ curves for interesting variations.  Some of the
results have been reported elsewhere (Rappaport et al.~2013; Tran et al.~2013).  The $O-C$
curve for KIC 9164561 is discussed in Sect.~\ref{sec:third}.  We 
also searched for folded light curves that (i) exhibit a flat-bottomed 
eclipse that is deeper than the other eclipse, and (ii) are associated with hot {\it Kepler} targets, i.e., targets that have a reported $T_{\rm eff} \gtrsim
7000$ K.  The deep flat-bottomed eclipse - high temperature combination is a signature of a hot white dwarf in orbit with a normal, F, A, 
or B star.  In addition to the four known {\em Kepler} low-mass bloated hot white dwarfs,
we found two others, previously unreported, in the binaries KIC 9164561 and KIC 10727668.

The nondegenerate components of KIC 9164561 and KIC 10727668 are, according to the Kepler
Input Catalog, hot with effective temperatures of $\sim$8000 K and 10,000 K, respectively, and thus are not susceptible to the 
starspot modulations that are exhibited by cooler stars.  An examination of the Fourier transform 
of the KIC 10727668 flux time series shows no evidence for any periodicities, including those from stellar pulsations, other 
than that of the orbit.  In the case of KIC 9164561, the Fourier 
transform of the flux time series shows a set of pulsations at frequencies $\gtrsim 20$ cycles/day,
which does not significantly affect our light curve or timing analysis.  We therefore made no 
corrections to the data except for normalizing each quarter to its median value.  

The orbital periods given in the {\em Kepler} binary catalog (Matijevic et al.~2012; 
Kirk et al.~2015) were used to produce initial folds of the 
data.  From these folds, we established a trial template for the orbital light curve.  This
profile was then used to fit each individual eclipse in the time series, thereby determining
a set of 893 and 108 eclipse times for KIC 9164561 and 10727668, respectively.  We then
fit a linear function to the sequence of eclipse times of the form: $t_n = t_0+nP_{\rm orb}$,
where $n$ is the nth eclipse.  The two orbital periods thus determined are in good 
agreement with those given in the {\em Kepler} binary catalog, but have somewhat
smaller uncertainties.  The periods and their uncertainties for these two systems are 
listed in Table \ref{tab:sys}.

We also fit the eclipse times for each system including a quadratic term to allow for possible 
changes in the orbital period or light travel time effects that might occur if the binary is part of a triple system.
These are discussed in Sect.~\ref{sec:third} below.

\begin{figure}
\begin{center}
\includegraphics[width=0.496 \textwidth]{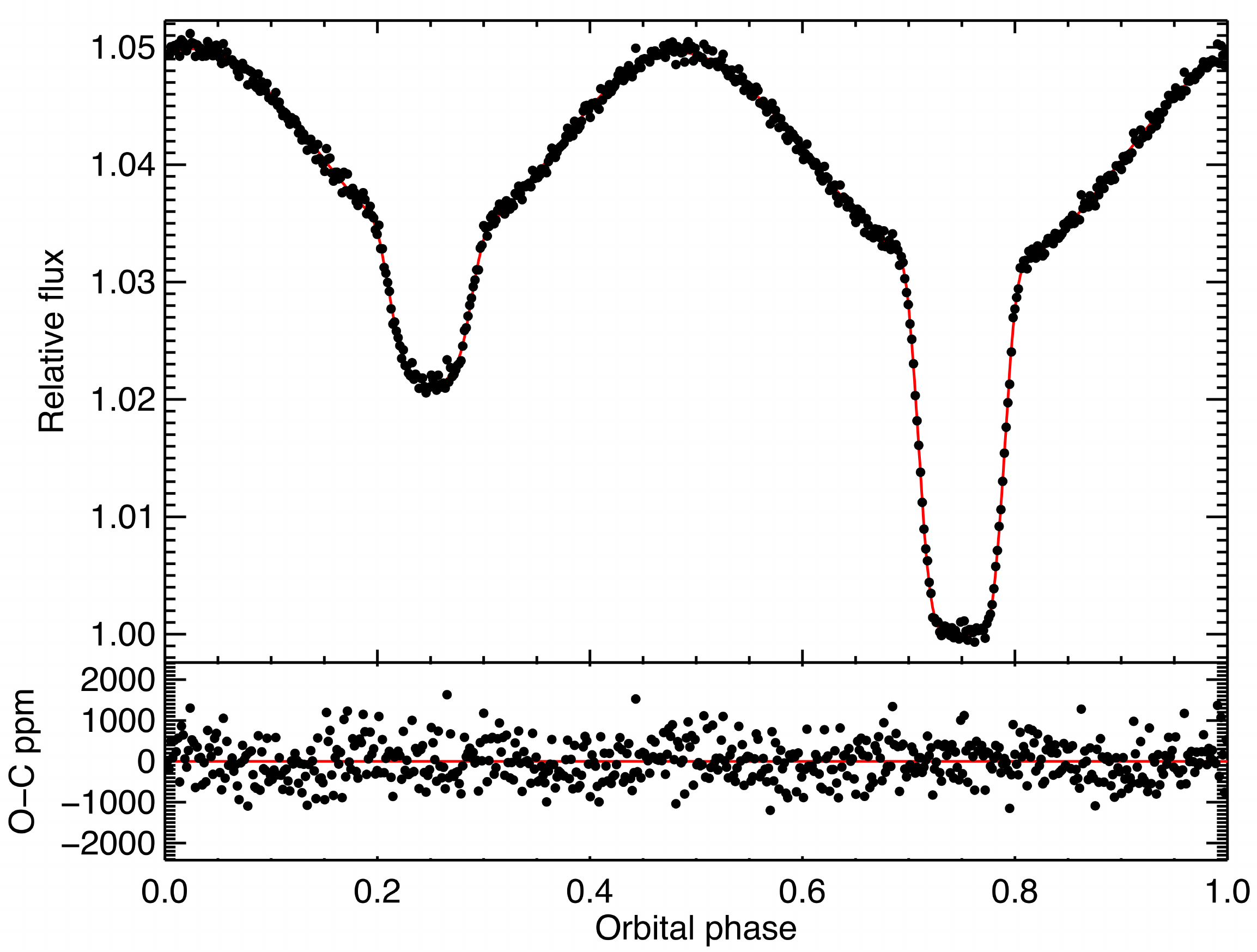} 
\caption{Folded light curve for KIC 9164561 for an orbital period of 1.2670400 days.
The folded data have been averaged into 3-minute bins.  The red curve is a fit to a 
model that is described in Sect.~\ref{sec:model}.  The bottom panel shows the
residuals of the data minus the fitted model.}
\label{fig:916}
\end{center}
\end{figure}

\begin{figure}
\begin{center}
\includegraphics[width=0.496 \textwidth]{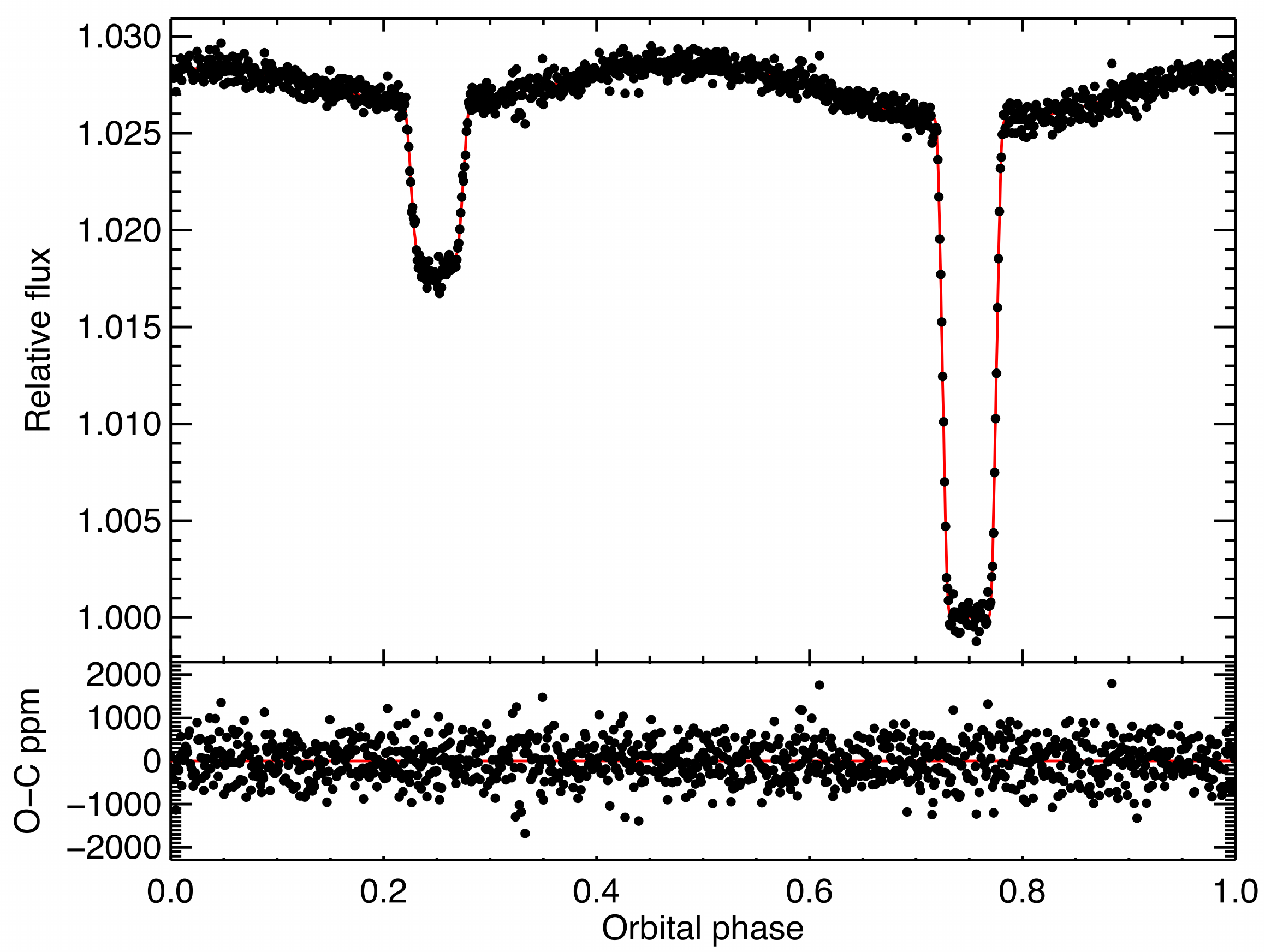}
\caption{Folded light curve for KIC 10727668 for an orbital period of 2.305897 days.
The folded data have been averaged into 3-minute bins.  The red curve is a fit to a 
model that is described in Sect.~\ref{sec:model}.  The bottom panel shows the
residuals of the data minus the fitted model. }
\label{fig:107}
\end{center}
\end{figure}

\vspace{0.6cm}

\section{Orbital Light Curves}
\label{sec:LCs}

The data for both KIC 9164561 and KIC 10727668 were folded at the
orbital periods we determined from fitting the individual eclipses, as described above.
The results were averaged into 3-minute bins, and are shown
in Figures \ref{fig:916} and \ref{fig:107}.

Due to the small curvature observed in the eclipse timing residuals for KIC 9164561 
(see Sect.~\ref{sec:third}) we removed a quadratic term from the eclipse phases 
before producing the folded light curve in Fig.~\ref{fig:916}. The phase corrections
amount to less than 0.002 days (or 0.0016 orbital cycles).

In each case, the light curves are normalized to unity at the bottom of the 
deeper eclipse when the hot white dwarf has been totally eclipsed.  This 
is then the light level of the primary A star at a minimum of its ellipsoidal variations.  
The relatively small separation of the two components of
KIC 9164561 is manifest by the short orbital period, the large amplitude of the ellipsoidal light variations
(``ELV''), and the duration of the eclipse relative to the orbital period.  

\begin{figure}
\begin{center}
\includegraphics[width=0.5 \textwidth]{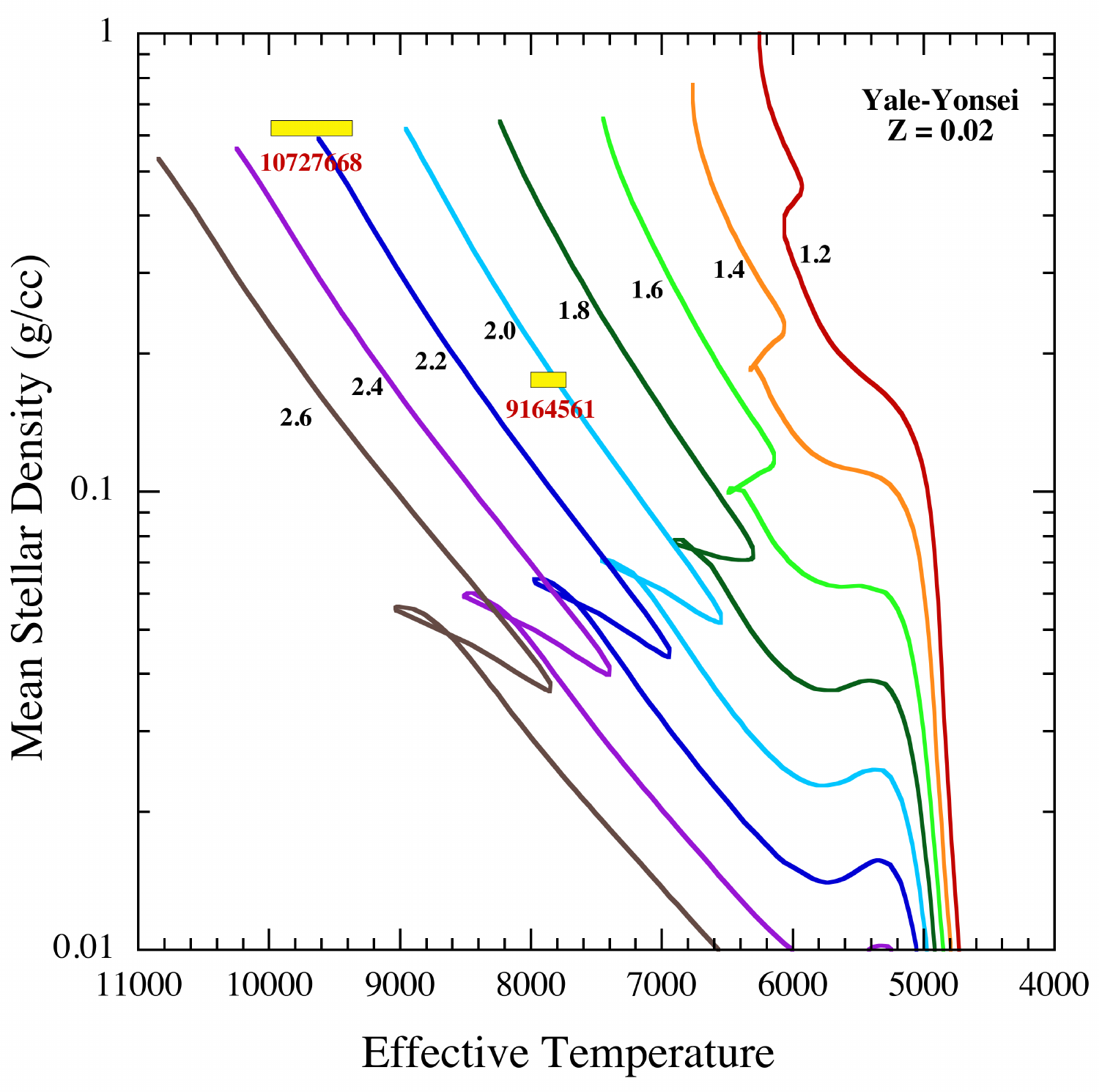} 
\caption{Stellar evolution tracks in the mean stellar density, $\langle \rho \rangle$, and $T_{\rm eff}$
plane.  The stellar masses range from 1.2 to 2.6 $M_\odot$ in steps of 0.2 $M_\odot$.  The tracks were 
produced with Yonsei-Yale stellar isochrones (Yi et al.~2001).  The locations of 
KIC 9164561 and KIC 10727668 are indicated by yellow rectangles (1 $\sigma$ confidence regions).  The mean stellar densities
are determined from the eclipse parameters.  }
\label{fig:YY}
\end{center}
\end{figure}

\subsection{Simple Geometrical Model for the Light Curves}
\label{sec:model}

The light curves of these two binaries were analyzed with a relatively simple
geometrical model which involves 13 free parameters.  In brief, there are 7 free parameters
associated with the eclipses: $R_{\rm wd}/R_1$, $R_1/a$, $i$, $L_{\rm wd}/L_1$,
two limb-darkening coefficients, and $t_0$, where subscript ``1'' refers to the
primary A star, $a$ is the orbital separation, $i$ the orbital inclination angle, and 
$t_0$, the time of the middle of a primary eclipse.  The out-of-eclipse phases
are fitted with a set of 6 terms, i.e., a sine and a cosine term for each of the orbital frequency and its next
two higher harmonics.  Each of these can be associated with different contributions
to the ELVs, the illumination term, and the Doppler boosting effect, as described
in some detail in Carter et al.~(2011).  

Once the best-fitting parameters were found with a Levenberg-Marquardt fitting
algorithm, the allowed region in parameter space was defined with a Markov
Chain Monte Carlo (MCMC) routine, which used the Gibbs sampler and the 
Metropolis-Hastings algorithm, and a likelihood proportional to $\exp(-\chi^2/2)$
(see, e.g., Holman et al.~2006; Sanchis-Ojeda et al.~2013). 

The results for the fitted model are plotted as red curves in Figs.~\ref{fig:916}
and \ref{fig:107}.  The high quality of the fit in each case is indicated by the white-noise 
appearance of the residuals.  The results for the fitted parameters and their uncertainties
are summarized in Table \ref{tab:sys}.  

\section{The A Star Companions}
\label{sec:Astars}

The mass and radius of the primary star in each
system are essential to quantitative models of the two binaries.  There are two accessible 
parameters that can be utilized to estimate the primary mass: the effective temperature, $T_1$, 
and the mean stellar density, $\bar{\rho}_1$.  The {\em Kepler} 
Input Catalog (KIC) gives $T_1 = \,\, $8059 \,K for the primary star of KIC 9164561 and 9696 K for the primary of KIC 10727668.
These temperatures are based on 5-color photometry.
For KIC 9164561 we measured $T_{\rm eff} = 7870 \pm 120$ K from the same spectra used for the radial velocity measurements (see Sect.~\ref{sec:spec}).  In the case of KIC 10727668, where we have no additional spectral information, we estimate the uncertainty in the KIC value of $T_1$ at approximately $\pm$ 300 K (Huber et al.~2104).  

The second readily, and more accurately, determined parameter associated with each primary star 
is its mean density, $\bar{\rho}_1 \equiv M_1/(4 \pi R_1^3/3)$.  This mean density is 
a function only of the light curve-determined parameters $P$ and $a/R_1$, with a weak 
dependence on $q$ (the mass ratio $M_2/M_1$).  The dependence of $\bar{\rho}_1$ on these parameters may be found by 
dividing Kepler's third law by $R_1^3$ to obtain:
\begin{eqnarray}
	\left(\frac{a}{R_1}\right)^3 & = & \frac{1}{3 \pi} G P^2 \bar{\rho}_1 (1+q) ~~.
\end{eqnarray}
This relation was pointed out by Seager \& Mall{\'e}n-Ornelas (2003) in the context of transiting 
exoplanets.   For exoplanets, the value of $q$ is much smaller than the error in $a/R_1$ so that 
$\bar{\rho}_1$ can be determined independently of any mass information.  While the same is not quite 
true with our two binaries, the value of $q$ is still sufficiently small, i.e., $q \lesssim 0.1$
that a rather accurate value of $\bar{\rho}_1$ may be determined nearly independently of $q$.  
In particular, by using Eqn.~(3) with the results from our photometric analysis of $R_1/a$ and with, as a first approximation, $q=0.1$, we find that $\bar{\rho}_1 \approx 0.17 \pm0.04$ g cm$^{-3}$ for KIC 9164561, and $\bar{\rho}_1 \simeq 0.63 \pm0.04$ g cm$^{-3}$ for KIC 10727668.  These values are refined below using estimates for $q$ that are based on estimates of the masses of the two component stars.

Figure \ref{fig:YY} shows stellar evolution tracks for stars of a range of initial masses in the $\bar{\rho}$--$T_{\rm eff}$ plane.
The locations of the two target primary stars are indicated by the rectangular boxes.  We conclude
from this that the mass of the primary star in KIC 9164561 must be  $M_1 = 2.02 \pm 0.06 \, M_\odot$ (KIC 9164561)
and that of the primary in KIC 10727668 must be $2.22 \,\pm\,0.10$ $M_\odot$.

\section{RV measurement of KIC 9164561}
\label{sec:RV}

The spectroscopic follow-up of KIC9164561 was carried out with the FIbre-fed \'Echelle Spectrograph \citep[FIES;][]{Frandsen1999,Telting2014} mounted at the 2.56-m Nordic Optical Telescope (NOT) of Roque de los Muchachos Observatory (La Palma, Spain). Spectra were taken under clear and stable sky conditions at 11 epochs between May and September 2014. We used the $1.3\,\arcsec$ \emph{med-res} fibre, which provides a resolving power of about 47000 in the spectral range 3600\,--\,7400\,\AA. The observations taken at each epoch were split into two or three consecutive sub-exposures of 1200 seconds each to remove cosmic ray hits. Following the observing strategy described in \citet{Buchhave2010}, we monitored the radial velocity drift of the instrument by acquiring long-exposed (T$_\mathrm{exp}$=15-30 sec) ThAr spectra in ``sandwich mode", i.e., right before and after each epoch observation. The data were reduced using a customized IDL software suite, which includes bias subtraction, flat fielding, order tracing and extraction, and wavelength calibration. Radial velocities (RVs) were determined by cross-correlating the target spectra with the spectrum of Vega, for which we adopted a heliocentric radial velocity of $-13.5$\,\kms \citep{AllendePrieto2004}.

\begin{figure}
\begin{center}
\includegraphics[width=0.99 \columnwidth]{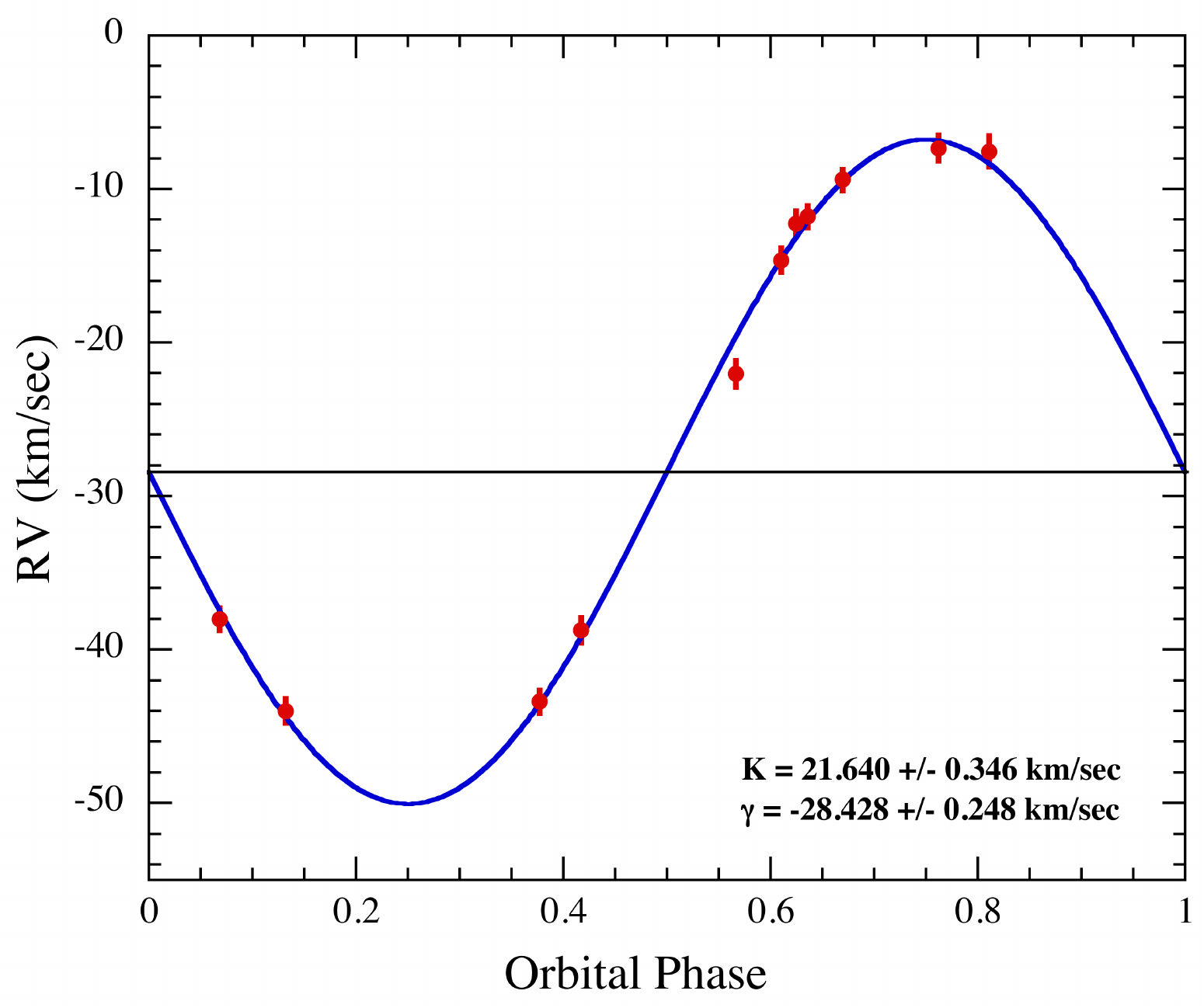} 
\caption{Radial velocity measurements (points with error bars) made
with the {\em FIES-NOT} spectrometer during the period 2014 June 15
through 2014 September 15.  The uncertainty per point is estimated to
be $\approx$ 700 m s$^{-1}$.  The curve is the best-fit circular orbit
with orbital period and phase taken from the {\em Kepler}
photometry. The K-velocity of the A star is 21.64 $\pm \, 0.35$ km
$^{-1}$, and the $\gamma$ velocity is $-28.43$ km$^{-1}$.}
\label{fig:RVs}
\end{center}
\end{figure}

Table ~\ref{RV-Table} lists the FIES RV measurements, the signal-to-noise (S/N) ratio per pixel at 4650 ~\AA, and the full-width at half maximum (FWHM) of the FIES cross-correlation function (CCF). Full-widths at half maximum were extracted by fitting a Gaussian to the CCFs between the observed spectra and the ATLAS9 model spectrum having the same photospheric parameters as the target star (see Sect.\,\ref{sec:spec}) and smoothed to the FIES spectral resolution.  Figure \ref{fig:RVs} shows the phase-folded RV curve of KIC9164561, based on the phase established from the {\em Kepler} photometry, and the Keplerian circular-orbit fit to the results.  The K velocity of the primary A star is 21.64 $\pm$ 0.35 km s$^{-1}$, and the corresponding $\gamma$-velocity is -28.43 $\pm$ 0.25 km s$^{-1}$. 

The corresponding mass function is 
$$f(M_2) = 0.00133 \pm 0.00006 ~M_\odot$$
For an orbital inclinational angle of $71.6^\circ \pm 0.2^\circ$ (see Table \ref{tab:sys}), this implies:
$$\frac{M_2^3}{(M_1+M_2)^2} = 0.00155 \pm 0.00008 ~M_\odot$$
The mass for the A star of $M_1 = 2.02 \pm 0.06$\,M$_\odot$ obtained above may be combined with the mass function to obtain a white dwarf
mass of $M_2 = 0.197 \pm 0.005 \,M_\odot$.  The mass ratio for the KIC
9164561 system is therefore $q = 0.097 \pm 0.004$.

Because KIC 10727668 is faint, i.e., 2.6 {\em Kepler} magnitudes fainter than KIC 9164561, we were unable to obtain any useful RV measurements for that system.

\begin{table*}
\caption{FIES radial velocity measurements of KIC 9164561}
\begin{center}
\label{RV-Table}
\begin{tabular}{ccccccc}
\hline
\hline
\noalign{\smallskip}
BJD$_{\rm UTC}$              &   RV    & $\sigma_{\mathrm RV}$ &  CCF$^a$ FWHM & $\sigma_{\rm FWHM}$ & S/N/pixel$^b$ & Date of Obser.$^c$   \\
($-$ 2\,450\,000)&   \kms  &    \kms     & \kms &   \kms       &  @4650\,\AA & (at start)  \\
\noalign{\smallskip}
\hline
\noalign{\smallskip}
6783.67056  &   -9.397   & 0.673 & 103.6 & 1.4 & 23 & 06 May 2014 \\
6833.58987   & -38.015   & 0.714 & 106.7 &1.2 &19 &  25 Jun 2014 \\
6833.67106   & -43.987 &  0.768 & 102.9 & 1.2 &17 &  25 Jun 2014 \\
6844.66674   &  -7.539 &  0.968 & 112.1 & 1.2 &11 & 06 Jul 2014 \\
6845.43519   & -38.741 &  0.794 & 109.1 & 1.0 &15 &  06 Jul 2014 \\
6845.71215  &  -11.787 &  0.690 & 102.8 & 1.2 &23 &  07 Jul 2014 \\
6901.37427  &  -22.004  &  0.847 & 107.6 & 1.2 &15 &  31 Aug 2014 \\
6902.40137  &  -43.375 &  0.743 & 104.6 & 1.3 &17 &  01 Sep 2014 \\
6915.38485  &  -12.215 &  0.784 & 106.5 & 1.3 &15 & 14 Sep 2014 \\
6915.55901   &  -7.337 &  0.820 & 107.2 & 1.1 &14 & 15 Sep 2014 \\
6920.43515  &  -14.633  & 0.791 & 105.6 & 1.3 &13 & 19 Sep 2014 \\
\noalign{\smallskip}
\hline
\end{tabular}
\end{center}
(a) Width of the cross correlation function. (b) S/N ratio per pixel at 4650\,\AA.  (c) Calendar date in UT at the beginning of the target observation.
\end{table*} 

\subsection{Spectral analysis of KIC 9164561}
\label{sec:spec}

\begin{figure}
 \begin{center}
\includegraphics[width=0.99 \columnwidth]{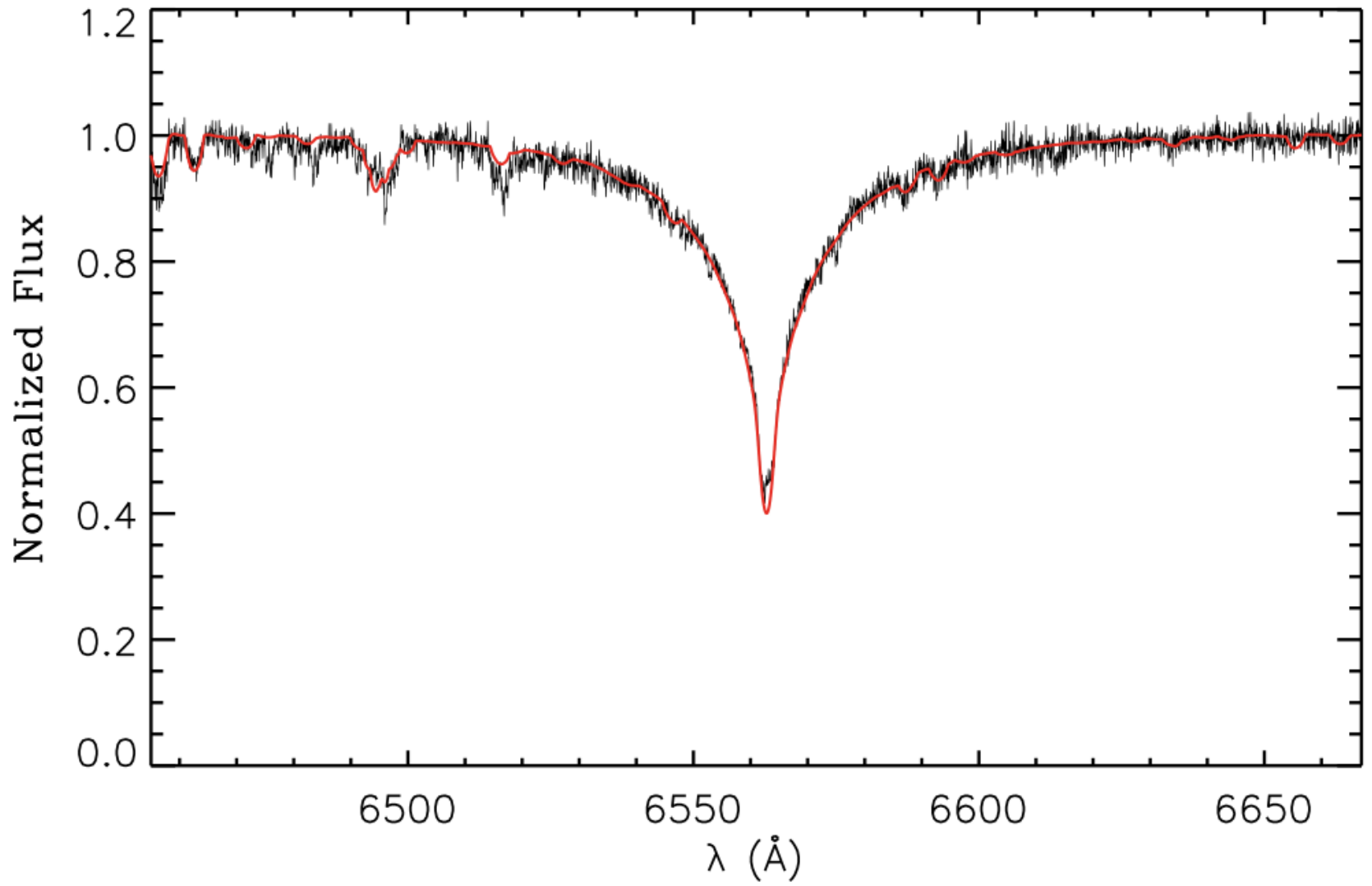} 
\includegraphics[width=0.99 \columnwidth]{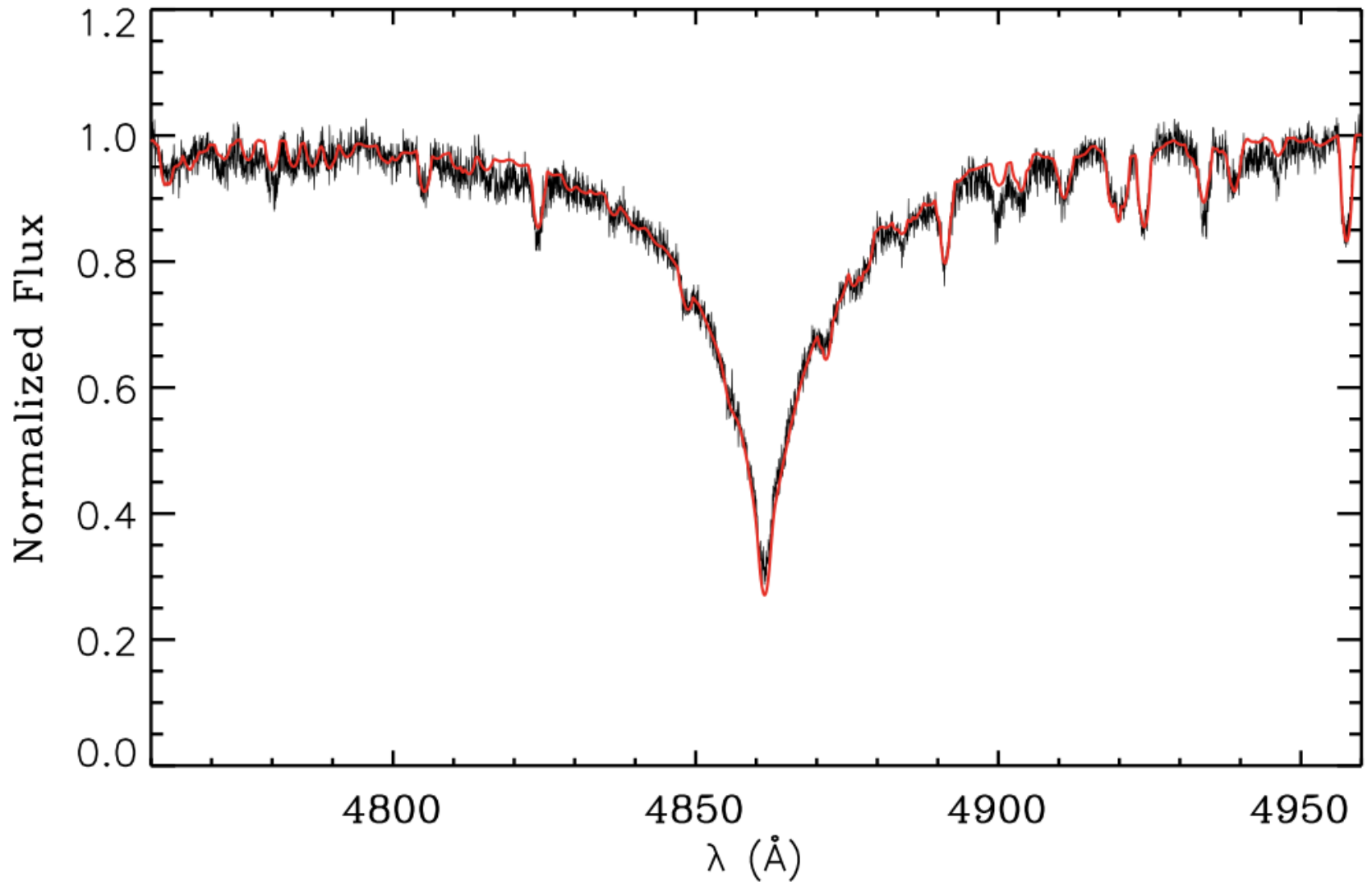}  
\includegraphics[width=0.99 \columnwidth]{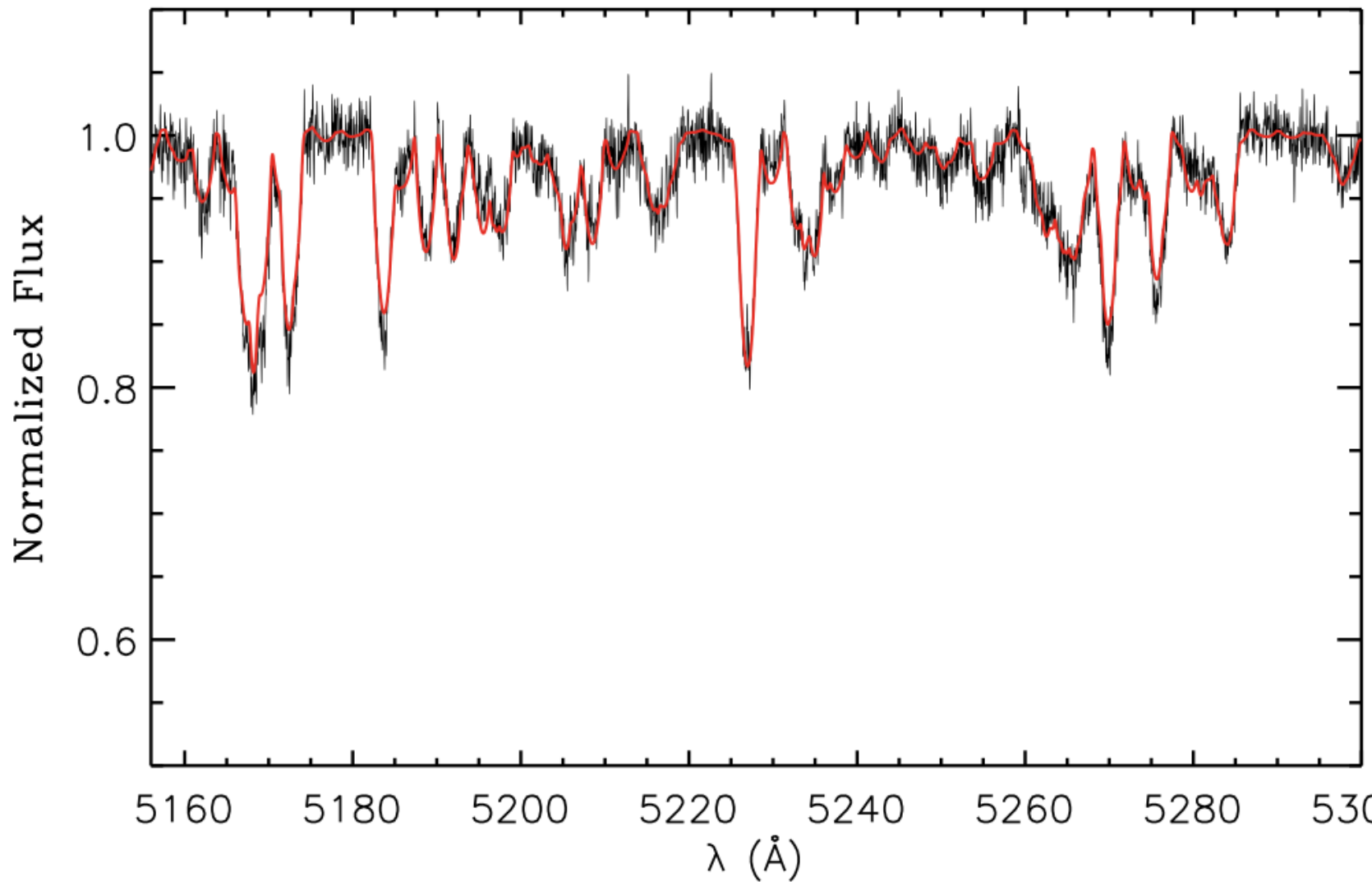} 
 \caption{FIES co-added spectrum of KIC\,9164561 (black line) encompassing the H$_{\alpha}$ (upper panel), H$_{\beta}$ (middle panel), and Mg\,{\sc i} b (lower panel) lines. The best fitting ATLAS9 model spectrum is overplotted with a thick red line.}
 \label{FIES_spectrum}
 \end{center}
 \end{figure}

We started the spectral analysis by coadding all the FIES spectra, after shifting them to a zero-velocity reference frame, which yielded a S/N of $\approx$ 55 per pixel at 4650\,\AA. The spectral parameters of the primary component in KIC 9164561 were then derived by fitting this spectrum to a grid of synthetic spectra calculated with the SPECTRUM code \citep{Gray1994} using ATLAS9 models \citep{Castelli2004}.
We assumed a secondary-to-primary flux ratio of 0.0285, as estimated from the eclipse light curve modeling (Table~\ref{tab:sys}). We adopted the calibration equations from \citet{AllendePrieto2004} and \citet{Bruntt2010} for the purpose of estimating the microturbulent $v_{\rm micro}$ and macroturbulent $v_{\rm macro}$ velocities. These equations were extrapolated to the effective temperature of the star. 
The spectral type was determined following the method described in \citet{Gandolfi2008}. The estimated stellar parameters are listed in Table \ref{Star-parm}. Figure~\ref{FIES_spectrum} shows the co-added FIES spectrum in the spectral region encompassing the H$_{\alpha}$, H$_{\beta}$, and Mg\,{\sc i} b lines, along with the best fitting synthetic model.

\begin{table}[t]
\centering
\caption{Spectroscopic parameters of the primary component of KIC\,9164561}
\label{Star-parm}
\begin{tabular}{lc}
            \hline
            \hline
            \noalign{\smallskip}
            Parameter      &  \text{Value} \\
            \noalign{\smallskip}
            \hline
            \noalign{\smallskip}
            Effective Temperature $T_\mathrm{eff}$ (K)                           & $7870\pm120$             \\
            Spectroscopic surface gravity log\,$g$ (log$_{10}$ cm\,s$^{-2}$)     & $4.0\pm0.1$              \\
            Metallicity [M/H] (dex)                                              & $0.0\pm0.1$              \\
            Microturbulent velocity $v_ {\mathrm{micro}}$ (km\,s$^{-1}$)         & $ 2.6\pm0.3$             \\
            Macroturbulent velocity $v_ {\mathrm{macro}}$ (km\,s$^{-1}$)         & $11.3\pm0.7$             \\
            Projected rotational velocity $v$\,sin\,$i$   (km\,s$^{-1}$)         & $  75\pm5$               \\
            Spectral Type                                                        & \text{A}7\,\text{V}      \\
            \noalign{\smallskip}
            \hline
\end{tabular}
\end{table}

 \subsection{Constraints on a Third Star in KIC 9164561}
 \label{sec:CCF}
 
 We looked for the spectroscopic signature of a third (bound) component to KIC 9164561 by cross-correlating the FIES spectra with a set of A- through M-type observed templates. The templates were gathered between 2011 and 2014, as part of the on-going RV follow-up of {\em CoRoT} and {\em Kepler} stars carried out with FIES@NOT (e.g., Gandolfi et al. 2013). 

Assuming that the third star is on or near the ZAMS, its mass cannot exceed $\sim$2 $M_\odot$ -- corresponding to an A7V spectral type -- without significantly violating a number of observational constraints (see below). The eclipse timing variations detected in the {\em Kepler} light curve (see Sect.~\ref{sec:third}) constrain the orbital period of a putative third star in the system to lie between 8 and 19~years, implying that its RV is within $\sim$12 km s$^{-1}$ of the $\gamma$ velocity of the binary system. The FIES RV follow-up covers a time span of about 4.5 months, which corresponds to less than 4.6\,\% of the orbital period of the third component, constraining its RV variation to be less than $\sim$4 km s$^{-1}$ within the observing window. Because of the relatively high projected rotational velocity of the primary star (see Table~\ref{Star-parm}), the almost ``stationary'' CCF peak of the third source is expected to be blended with the broad peak of the A7\,V component. 

The FIES CCFs show neither a peak with constant RV superimposed on top of the broad peak of the A7\,V star, nor a skewed profile that would suggest the presence of a another rapidly rotating star\footnote{The white dwarf is undetected in our spectroscopic data-set, owing to the small secondary-to-primary flux ratio of 0.0285 (see Table \ref{tab:sys}).}. In order to assess the significance of the non-detection and constrain the parameters of a possible third body, we built up a set of synthetic double-lined spectroscopic binary (`SB2') spectra using ATLAS9 model atmospheres. We adopted Gray's (2005) calibration for dwarf stars to convert spectral types to stellar parameters. We combined the rotationally broadened model spectrum of an A7\,V star -- resembling the primary component -- with that of dwarf stars with different spectral types (from A7\,V to M5\,V) and values of $v$\,sin\,$i$. We added random noise -- to account for the average S/N ratio of the FIES spectra -- and the contaminant light of the second source -- to account for the white dwarf. We assumed a zero velocity for the third light and radial velocities in the range $-21.64<$\,RV\,$<21.64$~km\,s$^{-1}$ for the A7\,V primary star. Finally, we cross correlated the combined SB2 spectra with narrow-lined synthetic spectra covering different spectra types, in the same fashion as for the observed spectra. 

Since the contrast of a CCF peak decreases with increasing $v$\,sin\,$i$, we considered the following two cases.

\begin{figure}[h]
   \centering
   \includegraphics[width=\linewidth]{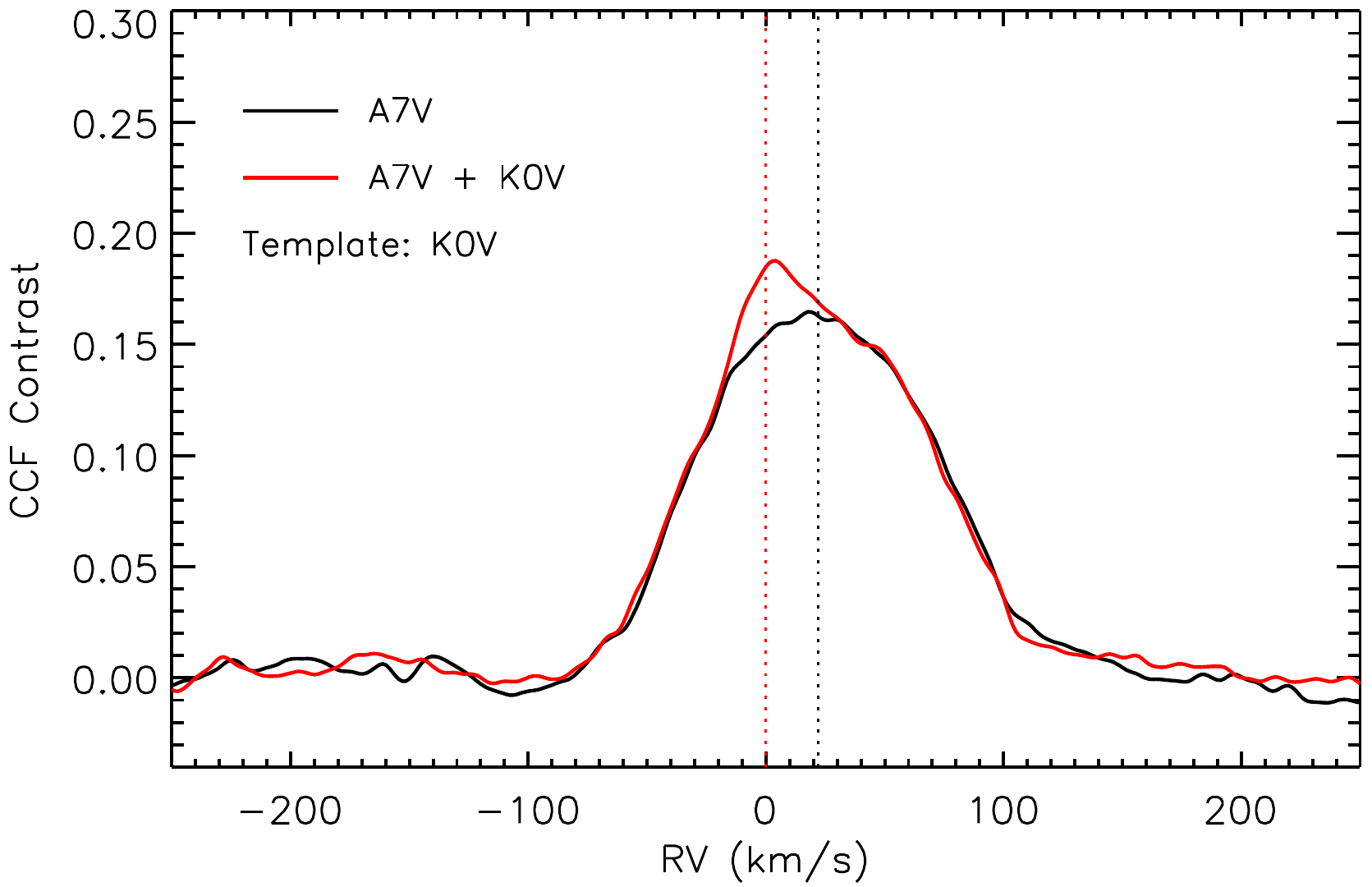}
   \caption{Simulated cross-correlation function (red curve) between a narrow-lined K0\,V synthetic template and the spectrum of an hypothetical double-lined spectroscopic binary containing an A7\,V with $v$\,sin\,$i = 75$\,km/s and a K0\,V star with $v$\,sin\,$i = 18$\,km/s (see text for details). The dashed red and blue lines mark the RV of the K0\,V (0 km/s) and the A7\,V star (21.6 km/s), respectively. The black curve, shown for reference, is the same CCF, but without the presence of the K0\,V star.}
\label{fig:simCCFs}
 \end{figure}

{\it Spectral type later than F5\,V}. We used gyro-chronology to constrain the $v$\,sin\,$i$ of the third star. Given the age of the system ($\sim$925~Myr; see Sect.~\ref{sec:cool}), the rotation period of the third component is expected to be longer than about 4~days for spectral later than F5\,V (Barnes et al. 2007). This places an upper limit of 18 km s$^{-1}$ on $v$\,sin\,$i$. Although this value likely overestimates the true $v$\,sin\,$i$ for the late-type stars, we conservatively assumed 18 km s$^{-1}$ to be the projected rotational velocity associated with the third light. We found that any stars with spectral type between K2\,V and F5\,V would be detectable in the FIES CCF (for an example see Fig.~\ref{fig:simCCFs}). For later spectral types the peak of the third light is embedded in the noise of the CCF, owing to the relatively low S/N ratio of the spectra and the low flux of the third star. 

{\it Spectral type between F5\,V and A7\,V}. For $v$\,sin\,$i<60$~km s$^{-1}$ the CCF peak of the third component is clearly visible on top of the peak of the primary star, as described in the previous case. For $v$\,sin\,$i>90$~km/s the CCF would assume a variable skewed profile, whose shape changes with the changing RV of the primary star, which we do not observe. Finally, for $60 ~{\rm km/s}<v$\,sin\,$i<90$~km/s the two CCF peaks are merged together and larger actual RV variations of the primary component would be needed to account for the measured RVs (see below).  In the case of comparable amounts of light contributed by the A7\,V star and a putative third star, we can say two things.  First, the FWHM of the CCF would vary in a correlated way with the measured RVs.  The extra widening can be estimated analytically, and the broadening factor is given by $[1+1/2 (\delta v/\sigma)^2]$, where $\delta v$ is the observed (i.e., `diluted') instantaneous value of the A7\,V star RV, and $\sigma$ is the Gaussian width of the CCF curve.  Since the CCF FWHM remains constant at 106.5 km s$^{-1}$ to within an rms scatter of $\sim$3 km s$^{-1}$ (see Table \ref{RV-Table}), and a 42 km s$^{-1}$ shift (i.e., $\delta v = 21$ km s$^{-1}$) is expected to widen the CCF by 12 km s$^{-1}$, which is distinctly not observed, we conclude that there is no other A- to early F-type star of comparable brightness in the system.  Second, as mentioned above, a stationary line would `dilute' the actual Doppler shift by a factor of $\sim$$\mathcal{F}_{\rm A7V}/(\mathcal{F}_{\rm A7V}+\mathcal{F}_{\rm 3})$, where  $\mathcal{F}_{\rm A7V}$ and $\mathcal{F}_{\rm 3}$ are the fluxes of the A7V star and the third body, respectively.  In this case, the properly inferred mass of the white dwarf (based on the RV amplitude) would be a factor of approximately $(1+\mathcal{F}_{\rm 3}/\mathcal{F}_{\rm A7V})$ higher.  This, in turn, would yield an implausibly high mass for the white dwarf when $\mathcal{F}_{\rm 3}$ exceeds $\mathcal{F}_{\rm A7V}$ by even a factor of 2.  Therefore, the mass of an early-type third body is independently constrained to be $\lesssim 2.5 \, M_\odot$.

Thus, based on our simulations, we can exclude nearly any plausible near-MS third star that has a spectral type earlier than K2\,V.  While there is still some small `wiggle room' left in parameter space, these constraints likely imply an upper limit for the mass of any third star to be about 0.8~$M_\odot$.

\section{Doppler Boosting Effect}
\label{sec:DB}

In principle, the Doppler boosting amplitude (Loeb \& Gaudi 2003; van Kerkwijk et al.~2010) can serve as a surrogate in place of radial velocity (`RV') measurements for measuring the binary K velocity.  As formulated by Carter et al.~(2011) the Doppler boosting (hereafter `DB')  amplitude in a binary is given by:
\begin{equation}
\label{eqn:db}
A_{\rm db} = \frac{K_1 \alpha_1}{c} \left[1-\frac{\alpha_2}{\alpha_1}\frac{F_2}{F_1}\frac{M_1}{M_2}\right]
\end{equation}
where the $F$'s are the fluxes of the two stars in the {\em Kepler} band.  The $\alpha$'s are Doppler boosting pre-factors calculated from:
\begin{equation}
\alpha = 3 - \left\langle \frac{d \ln F_\nu}{d \ln \nu}\right\rangle
\label{eqn:alpha}
\end{equation}
where $F_\nu$ is the spectral flux density at frequency $\nu$, and $\alpha$ must be evaluated using stellar atmosphere models 
averaged over the {\em Kepler} band.  The second term in Eqn.~(\ref{eqn:db}) represents the contribution to the DB amplitude from the white dwarf in the system (note the minus sign).  This method for evaluating the K velocities has been verified by Ehrenreich et al.~(2011) and Bloemen et al.~(2012) for the case of KOI 74, a system quite analogous to the ones discussed in this paper.

{\em KIC 9164561}: The measured Doppler boosting amplitude in this system, i.e., the amplitude of the $\sin \omega t$ term, where $\omega$ is the orbital frequency, and $t$ is time, is -10 $\pm$ 26 parts per million (`ppm').  Given the $K_1$ velocity obtained from the RV measurements (see Sect.~\ref{sec:RV}), for nominal values of $\alpha_1$, $A_{\rm db}$ for this system should be $\approx$160 ppm.  One possible explanation for this discrepancy is based on the large illumination semi-amplitude, i.e., the coefficient of the $\cos \omega t$ term, that is measured to be 2385 $\pm$ 36 ppm.  In this case, if the phase of the illumination is shifted in the proper sense from its nominal alignment with the eclipses by as little as 4$^\circ$, the entire DB effect would be cancelled. We note that in KOI 74 the illumination term is 600 times smaller because of the wider orbital separation and the lower luminosity of the white dwarf, and is therefore very unlikely to materially affect the observed Doppler boosting amplitude.  In order for the phase of the illumination effect in KIC 9164561 to be shifted by the amount required to cancel most of the Doppler boosting signal, the A-star primary would have to be rotating faster than the rate that is synchronous with the orbit, such that some of the heated surface material is carried into the night side of the star.  The measured rotation velocity of this star is $\approx$ 75 $\pm$ 5 km s$^{-1}$, a value that is to be compared with the equatorial rotation velocity of a star of radius 2.5 $R_\odot$ (see Table \ref{tab:sys}) rotating with a period of 1.267 days, i.e., $\approx$ 100 km s$^{-1}$.  Thus, if anything, the primary is rotating subsynchronously and the DB signal is therefore expected to be enhanced relative to its nominal amplitude. This is a potentially important discrepancy which merits follow-up study.

{\em KIC 10727668}: In this system, the Doppler boosting signal is measured to have a significantly non-negative value, 143 $\pm$ 20 km s$^{-1}$.  The illumination amplitude, 557 ppm, is 4 times smaller than in the case of KIC 9164561; this system may therefore be less susceptible to an illumination-induced perturbation of the DB amplitude.  If Eqn.~(\ref{eqn:db}) is used to evaluate $K_1$ we find:
\begin{equation}
K_1 \simeq \frac{20.0 \pm 2.8} {\left(1-0.02 M_1/M_2\right)}~{\rm km~s}^{-1}
\end{equation}
where we have taken $\alpha_1=2.15$, $\alpha_2 = 1.78$ (van Kerkwijk, private communication), and $F_2/F_1 \simeq 0.025$ (see Fig.~\ref{fig:107}).  This expression for $K_1$ may be entered into the mass function, and the cube root of the result taken to obtain the following equation involving the masses of the two stars:
\begin{equation}
\frac{M_2}{\left(M_1 + M_2\right)^{2/3}} \simeq \frac{0.107 \pm 0.015}{\left(1-0.02 M_1/M_2\right)} M_\odot^{1/3}
\end{equation}
where we have taken the orbital inclination angle to be 86.4$^\circ$ (see Table \ref{tab:sys}).  This equation can be solved numerically for $M_2$ given any specific value for $M_1$.  For a primary star mass of 2.2 $M_\odot$ (see Fig.~\ref{fig:YY}), the mass of the secondary is 
$$M_2 \simeq 0.266 \pm 0.035~M_\odot$$
where the error bar takes into account the uncertainty in $M_1$ as well as the uncertainty in $K_1$.  Again we caution that this estimate for $M_2$ does not take into account any systematic error due to an ``amplitude leak'' in the DB term.

\section{Ellipsoidal Light Variations}
\label{sec:ELV}

The ellipsoidal light variations (`ELVs') seen in the out-of-eclipse region of the light curves also contain information on the mass ratios, $ q = M_2/M_1$.  The basic expression for the ELVs (Kopal 1959; Morris \& Naftilan 1993; Carter et al.~2010) is:
\begin{equation}
B_{\rm ELV} \simeq -Z_1(2)\, q \,(R_1/a)^3 \sin^2 i
\label{eqn:elv}
\end{equation}
and 
\begin{equation}
Z_1(2) \simeq \frac{45 +3u}{20(3-u)}(1+\tau)
\end{equation}
where $B_{\rm ELV}$ is the amplitude of the cosine term at twice the orbital frequency, $u$ is the linear limb-darkening coefficient, and $\tau$ is the gravity darkening coefficient.  For the hot primary stars in KIC 9164561 and KIC 10727668 we take $u = 0.3$ and $\tau = 1$.  This leads to a numerical value for $Z_1(2)$ of 1.68.  We can then utilize the measured values of $B_{\rm ELV}$ of $-8880$ ppm and $-1179$ ppm for the two systems, respectively (see Table \ref{tab:sys}).  Similarly, we take the values for $R_1/a$ and $i$ for each system from Table \ref{tab:sys}. Finally, we solve for $q$ to find:
\begin{eqnarray}
q_{\rm ELV} (9164561) & = & 0.104 \pm  0.002\\
q_{\rm ELV} (10727668) & = & 0.15 \pm 0.01
\label{eqn:qelv}
\end{eqnarray}
where the error bars are statistical only and do not include any systematic uncertainties in the use of Eqn.~(\ref{eqn:elv}) to evaluate $q$ from the ELV amplitude.  

The value of $q$ found for KIC 9164561 from the RVs is $0.097 \pm 0.004$ (Sect.~\ref{sec:spec}) which is in quite satisfactory agreement with $q_{\rm ELV}$.

On the other hand, the value of $q$ for KIC 10727668 found from the ELV analysis (Eqn.~(\ref{eqn:qelv})), while still in marginal agreement with the value obtained from the Doppler boosting analysis ($0.12 \pm 0.02$), would imply a white dwarf mass as high as $M_{\rm wd} \simeq 0.33 \, M_\odot$.  This is quite on the high side of what we expect for this system (see Sect.~\ref{sec:discuss}).

\section{Masses and Radii of the White Dwarfs}
\label{sec:WDs}

In order to determine the sizes of the binary systems we need the total 
masses.  In the case of KIC 9164561, where a direct RV measurement has been made,
the total system mass is 2.22 $\pm$ 0.07 $M_\odot$.  With the application of Kepler's third
law, this leads immediately to an orbital separation of $a = 6.40 \, \pm 0.07 \,R_\odot$.  
For KIC 10727668 the Doppler boosting
analysis yielded a mass ratio $q \simeq 0.12$. If we tentatively accept this mass ratio,
and apply Kepler's third law, we find the orbital separation is 
$a = 9.95 \, \pm 0.15 \,R_\odot$.  The reason that the uncertainty is quite small,
in spite of the larger uncertainty in $q$ for this system, is that $a$ depends
only on the cube root of the total system mass, and $q$ is a small number. These
values for the orbital separation are summarized in Table \ref{tab:sys}.

The radii of the primary and secondary stars follow immediately 
from the $R_1/a$ and $R_2/a$ values derived from the light curve analyses and the values 
of the orbital separations.  These radii are summarized in Table \ref{tab:sys}.  Of particular note are 
the large radii of the two white dwarfs, 0.277 $R_\odot$ and 0.151 $R_\odot$ for KIC 9164561 and 
KIC 10727668, respectively, which indicate that these stars are quite thermally bloated.  The degenerate radii 
for cold white dwarfs of either He or hybrid HeCO compositions, are given by Eqn.~(4) in Nelson \& Rappaport
(2003).  The degenerate radii for the white dwarfs in KIC 9164561 and KIC 10727668 are 0.020 $R_\odot$
and 0.018 $R_\odot$, respectively, implying thermal bloating factors of 14 and 8.

\section{Evidence for a Third Star in KIC 9164561}
\label{sec:third}

\begin{figure}
\begin{center}
\includegraphics[width=0.47 \textwidth]{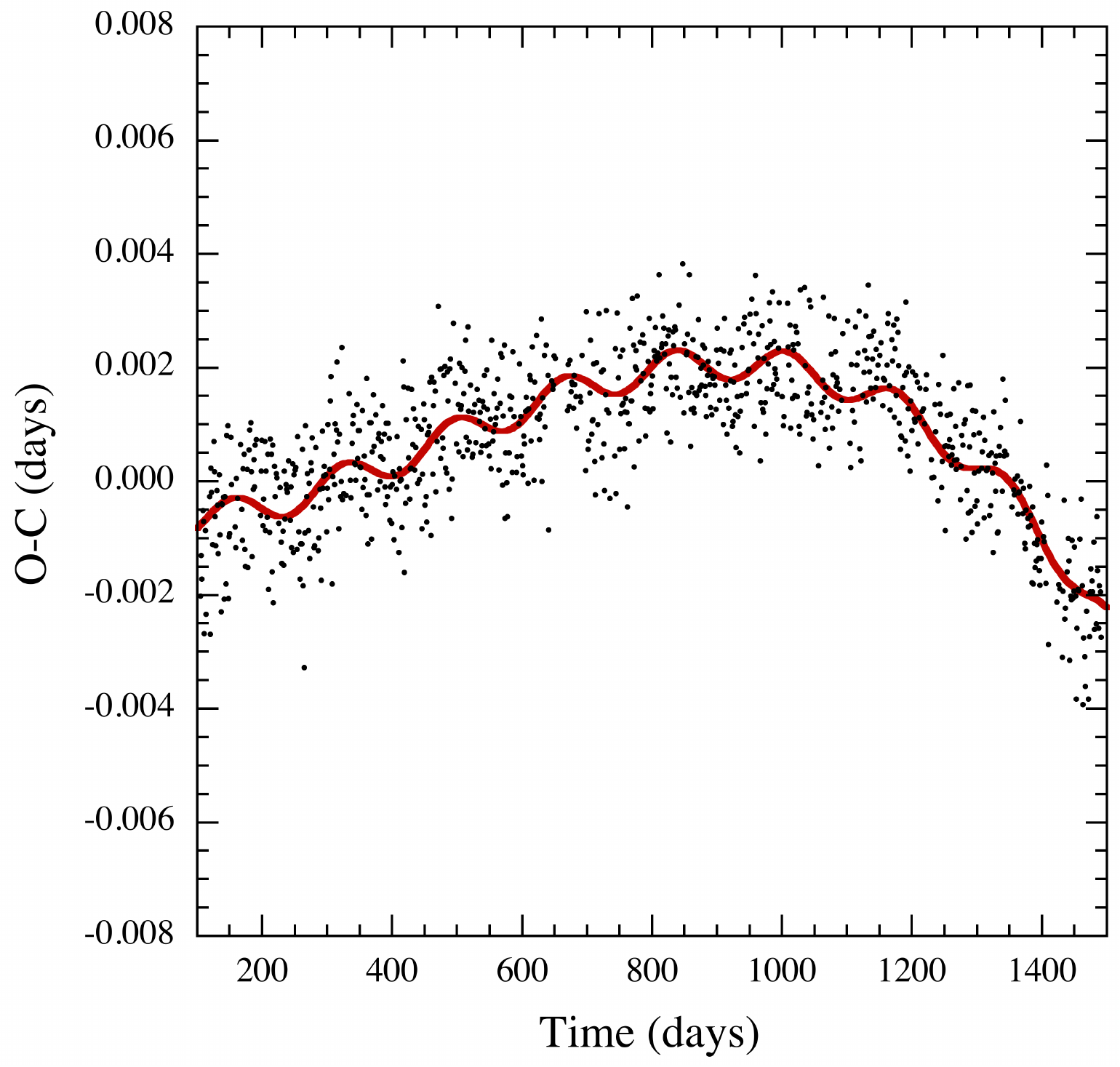} 
\caption{$O-C$ curve averaged over the primary and secondary eclipses of KIC 9164561 plotted as black points.  The $O-C$ curve is based on a binary period of 1.267040 days (see Table \ref{tab:sys}).  It is evident that the $O-C$ curve has a distinct curvature -- significant at the 35-$\sigma$ statistical level.  In addition, there is a small-amplitude oscillation with a predictable period of $\sim$170 days which is an artifact caused by a beat between the orbital period and the {\em Kepler} long-cadence sampling time. The characteristic timescale for the orbital period change implied by the overall long-term quadratic term is $\sim$$2 \times 10^5$ years.  If, instead, the curvature is attributed to a Roemer delay induced by a third body, the inferred orbital period is in the range of 8--19 years.  Here the red curve is the best-fitting circular orbit with a period of 13.7 years, and it includes a sinsuoidal modulation with a 170-day period to account for the artifact caused by the beats.}
\label{fig:OmC}
\end{center}
\end{figure}

We carried out an eclipse timing analysis for both binaries discussed
in this paper.  We used a code to produce the observed minus calculated curve (`$O-C$' curve)
that was tested in our search for triple star systems (Rappaport et al.~2013).  Only the $O-C$ 
curve for KIC 9164561 showed an interesting non-linear behavior.  The results are shown
in Fig.~\ref{fig:OmC} with black points for the average of the $O-C$ curves for the primary and secondary eclipses.  The $O-C$
curve shows two notable features: (i) a clear curvature, and (ii) small-amplitude
oscillations with a period of $\sim$170 days.  The small amplitude oscillations are due to
an artifact that results from a beat between the orbital period and {\em Kepler} long-cadence
sampling (see Eqns.~(4) and (5) in Rappaport et al.~2013).

\begin{figure}
\begin{center}
\includegraphics[width=0.47 \textwidth]{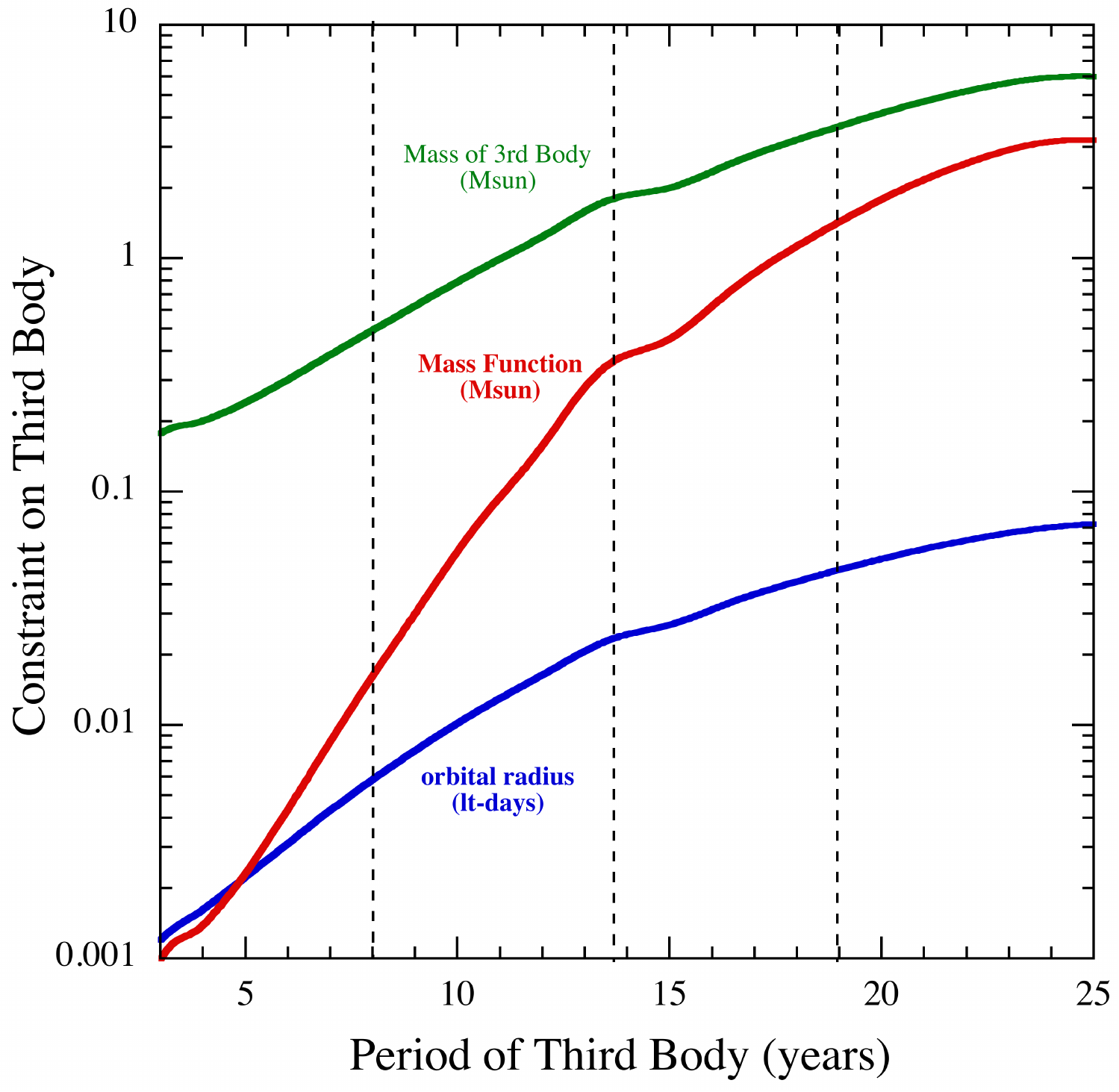} 
\caption{Illustrative constraints on the projected orbital radius (blue curve), mass function (red curve), and mass of the third body (green curve) in KIC 9614561 based on the $O-C$ curve shown in Fig.~\ref{fig:OmC}.  Only third-body orbital periods between 8 and 19 years yield somewhat acceptable fits (marked by dashed vertical lines), and periods greater than 14 years are excluded because the inferred third-body would be too massive and therefore likely too luminous to be missed in our spectroscopy.  Only circular coplanar orbits were considered in setting these constraints. }
\label{fig:constraints}
\end{center}
\end{figure}

The quadratic behavior in the $O-C$ curve is highly significant (35 $\sigma$), and can be 
interpreted as either a change in the orbital period, or as a Roemer delay due to the motion of
the binary induced by a third body.  The first option requires changes in the binary orbital period 
on a characteristic timescale of $P/\dot P \simeq 2 \times 10^5$ years.  Since there is no significant
mass transfer taking place in the system at the current time, this cannot be the cause of the 
apparent orbital period changes.  One would have to invoke another mechanism, such as 
the Applegate effect (Applegate \& Patterson 1987; Applegate 1992; Kotze \& Charles 2010) to explain the 
short timescale changes in the orbit.   We focus here on a more interesting and, in our view, more likely scenario, 
i.e., that there is a third body in the system.  It is then possible to constrain the orbital radius of such a third body 
as well as its mass, if we restrict our attention to coplanar and circular orbits.
Otherwise, the possibilities are nearly limitless.  

The results of such constraints are shown in Fig.~\ref{fig:constraints}.  We find that the orbital
period must lie in the range of $8 < P_{\rm trip} < 19$ years in order to provide at least a somewhat satisfactory fit to the 
$O-C$ curve ($\pm4\,\sigma$ limits).  The blue curve shows the best fit (projected) orbital radius as a function of $P_{\rm trip}$.  The red curve is the corresponding mass function, while the green curve is the mass of the third
body under the assumptions that the mass of the inner binary is $2.2 \,M_\odot$ and that the 
orbital inclination of the third body orbit is $90^\circ$.  Since the mass of the third body, if it is assumed to be on or near the ZAMS, cannot exceed $\sim$2 $M_\odot$ without affecting the spectrum, RV analysis, and the third light of the system in ways that are not observed (see Sects.~\ref{sec:spec} and \ref{sec:CCF}), the orbital period is further restricted to the range of $8 < P_{\rm trip} < 14$ years.  If we adopt the formal constraint on the mass of a main-sequence third star in the system from Sec.~\ref{sec:CCF} of $M_3 \lesssim 0.8 \, M_\odot$ we find that: $0.5 \lesssim M_3 \lesssim 0.8 \,M_\odot$ and $8 \lesssim P_3 \lesssim 10$ years.

To this point we have not yet considered the possibility that the putative third body might be a white dwarf.  The orbital period would then be restricted to the range $8-12$ years since the mass would have to be $< 1.4 \, M_\odot$ (see Fig.~\ref{fig:constraints}). If there had been a prior episode of mass transfer from the progenitor of the outer body white dwarf to the inner binary (KIC 9164561; see, e.g., Tauris \& van den Heuvel 2014) then, according to Eqn.~(\ref{eqn:Mwd_P}), the mass would lie in the range of $0.6-0.8\, M_\odot$.  Such white dwarfs would have cooled to the point where they would contribute very little light to the system ($\lesssim 10^{-4} L_{\rm tot}$).

Based on the inferred luminosity of KIC 9164561, and the appropriate bolometric corrections, we infer that its distance from the Earth is $\approx 390$ pc.  Typical orbital separations to the putative third body, for periods in the range of $8-19$ years, are expected to be $\approx 5-10$ AU.  At a distance of 390 pc, the corresponding angular separations are then $14-26$ mas.  This may be detectable with {\em Gaia}.

\begin{table}
\caption{KIC 9164561 and KIC 10727668 System Parameters}
\centering
\begin{tabular}{c c c}
\hline 
\hline
parameter & KIC 9164561 & KIC 10727668 \\  [0.5ex] 
%\startdata
\hline
RA (J2000) & 19h 42m 27.64s & 19h 21m 45.58s \\
Dec (J2000) & 45$^\circ$ 30$'$ 17.0$''$ & 48$^\circ$ 02$'$ 42.4$''$ \\
$P$ [days]  & 1.2670400 $\pm \,0.0000002$ & 2.305897 $\pm \,0.000007$ \\
$t_0$ [MJD]$^{\rm a}$ & 64.61956 $\pm \,0.00013$ & 1207.3677 $\pm \,0.0005$ \\
$K_p$ [mag.] & 13.71 & 16.32 \\
$R_2/R_1$ &  0.1091 $\pm \,0.0016$ & 0.0893 $\pm \, 0.0006$ \\
$R_1/a$ & 0.397 $\pm\, 0.003$ & 0.170 $\pm \, 0.003$  \\
$a$ [$R_\odot$]  & 6.40 $\pm \, 0.07$ & 9.95 $\pm$ 0.15 \\
$F_{K,2}/F_{K,1}$$^{\rm b}$ & 0.0285 $\pm \, 0.001$ & 0.0245 $\pm \, 0.001$ \\
inclin.~[deg] & 71.59 $\pm \,0.22$ & 86.45 $\pm \, 0.5$ \\
\\ \hline
$M_1$ $[M_\odot]$ & $2.02 \,\pm \, 0.06$ & $2.22\,\pm \,0.10$ \\
$R_1$ [$R_\odot$] & 2.54 $\pm$ 0.03 & 1.69 $\pm$ 0.04\\
$T_1$ [K] & 7870 $\pm$ 150 & 9696 $\pm$ 300 \\
$L_1$ [$L_\odot$] & 22.3 $\pm$ 1.5 & 22.9 $\pm$ 3.1 \\
$\rho_1$ [g/cc] & 0.171 $\pm$\,0.04 & 0.63 $\pm$ 0.04\\
$\log(g_1)$ [cgs] & 3.93 $\pm$ 0.02 & 4.32 $\pm$ 0.04\\
\\ \hline
$M_2$ [$M_\odot$] & 0.197 $\pm$ 0.005 & 0.266 $\pm$ 0.035 \\ 
$R_2$ [$R_\odot$] & 0.277 $\pm$ 0.005 & 0.151 $\pm$ 0.004\\
$T_2$ [K] & 10410 $\pm$ 200 & 14110 $\pm$ 440\\
$L_2$ [$L_\odot$] & 0.82 $\pm$ 0.07 & 0.82 $\pm$ 0.11\\
$\rho_2$ [g/cc] & 12.8 $\pm$ 1.2 & 120 $\pm$19 \\ 
$\log(g_2)$ [cgs] & 4.84 $\pm$ 0.04 & 5.55 $\pm$ 0.06 \\  
\hline
\multicolumn{3}{l}{Harmonic Amplitudes [ppm]}\\ \hline
$A_1$ [$\sin(\phi)$]$^{\rm c}$ & $-10$ $\pm$ 27 & 143 $\pm$ 20 \\
$B_1$ [$\cos(\phi)$]$^{\rm d}$ &  2385 $\pm$ 36 & 557 $\pm$ \,22\\
$A_2$ [$\sin(2\phi)$] & $-602$ $\pm$ 27 & $-47$ $\pm$ 20 \\
$B_2$ [$\cos(2\phi)$]$^{\rm e}$ & $-8880$ $\pm$ 35 & $-1179$ $\pm$ 21 \\
$A_3$ [$\sin(3\phi)$] &  $-30$ $\pm$ 28 & $-4$ $\pm$ 19 \\
$B_3$ [$\cos(3\phi)$] &  $-888$ $\pm$ 36 & $-48$ $\pm$ 23 \\
%\enddata
\hline
\end{tabular}
\label{tab:sys}
\tablecomments{a = BJD-2454900;  b = ratio of white dwarf flux to primary A-star flux in the {\em Kepler} band; c = Doppler boosting amplitude;   d = main illumination term;  e = main ellipsoidal light variation term.  See \S\,\ref{sec:DB} and \S\,\ref{sec:ELV} for details.}
\end{table}

\section{Discussion}
\label{sec:discuss}

Here we discuss the properties of the KIC 9164561 and KIC 10727668
systems and some of the evolutionary paths that may have led to their
formation.  We can then check whether the expected system parameters
match those determined empirically.

\subsection{Evolutionary Pathways}

There are two important evolutionary pathways that could lead to the formation of low-mass helium white dwarfs in close binaries (see, e.g., van Kerkwijk et al. 2010; Carter et al.~2011; Breton et al. 2012). The first channel includes one episode of common envelope (CE) evolution wherein the more massive star of the primordial binary overflows its Roche lobe after having become a giant with a degenerate core, and then undergoes dynamically unstable mass transfer onto its binary companion (see, e.g., Taam et al. 1978; Webbink 1984; Taam and Bodenheimer 1992). The rapid rate of mass transfer causes the companion to spiral inward towards the degenerate core of the giant while concomitantly expelling the envelope of the giant into the ISM. The resulting binary is composed of a degenerate white dwarf in a tight circular orbit with a largely unchanged companion star.

The second channel involves stable mass transfer between the components of the progenitor binary.  Unfortunately the degree to which mass transfer is conservative and the effectiveness of orbital angular momentum loss mechanisms (e.g., magnetic braking; see, for example, Verbunt \& Zwaan 1981) are uncertain. For reasonable physical assumptions in regard to the mass transfer and angular momentum loss processes, this channel can explain the formation and observed properties of these two systems (see Nangreaves et al.~2015, and references therein). 

{\em (i) The Common Envelope Channel}

Although a physically realistic 3-D hydrodynamic simulation of CE evolution has not yet been carried out, a reasonably good approximation of the effects of the process can presumably be obtained from first principles using energy conservation. As the companion star moves at high velocity through the envelope of the giant, a fraction $\xi$ of the orbital energy is transferred to the tenuous envelope material.  The companion star then spirals inward and more and more energy and angular momentum are transferred to the giant envelope. The gravitational potential energy that is released may be sufficient to eject the envelope. It is assumed that this process occurs on a timescale much shorter than the thermal timescale of the giant star. One of the more commonly used CE energy constraints was analyzed by Dewi \& Tauris (2000). It is given by 
\begin{equation}
\frac{G\left(M_c+M_e\right)M_e}{\lambda R_{1,0}}=\xi \left[\frac{GM_c M_{2,0}}{2 a_f}-\frac{G(M_c+M_e)M_{2,0}}{2a_0}\right]
\label{eqn:CE}
\end{equation}
where $M_c$, $M_e$, and $M_{2,0}$ are the core and envelope masses of the WD progenitor, and the mass of primordial secondary, respectively (see, e.g., Webbink 1984; Pfahl et al.~2003).  Note that $\xi$ is a dimensionless measure of the efficiency of the conversion of orbital energy that goes toward unbinding the envelope of the giant (the energy required to eject the envelope, $\Delta E_B$, must satisfy $\Delta E_B = \xi \Delta E_{\rm orb}$) and is usually taken to be $\lesssim 1$. The dimensionless factor $\lambda$ is intended to take into account the actual binding energy of the giant at the onset of the CE phase. It largely depends on the ratio of the core mass to the pre-CE-phase mass of the giant. Values of $\lambda$ have been tabulated for many different evolutionary states and masses of giants by Tauris \& Dewi (2001) and Podsiadlowski et al.~(2003b). For the types of giants that could produce helium white dwarfs, $\lambda$ is of order unity, normally having a value between 0.5 and 1.0\,\footnote{If we take $\lambda < 0.5$, which is possible in certain circumstances, this only makes the argument presented here more robust.}. Also note in equation (\ref{eqn:CE}) that $M_c$, the core mass of the giant, is approximately equal to the final white dwarf mass, $M_{\rm wd}$. Thus the original mass of the primary can be written as $M_{1,0} = M_c+M_e$ and its radius is given by $R_{1,0}$. Finally, $a_f$ is the separation of the white dwarf and secondary star in their mutual circular orbit after the CE phase while $a_0$ is the separation of the components of the progenitor binary. Using Kepler's third law, equation (\ref{eqn:CE}) can be rewritten as 
\begin{equation}
\frac{\eta}{2(q_0-q_f)}\left[\frac{q_f}{q_0}\left(\frac{1+q_0}{1+q_f}\right)^{1/3}\left(\frac{P_0}{P_f}\right)^{2/3}-1\right]\left(\frac{R_{1,0}}{a_0}\right)=1
\label{eqn:CE2}
\end{equation}
where $\eta = \xi \lambda$ is a dimensionless parameter that is $\lesssim 1$. Also $q_0 (\equiv M_{1,0}/M_{2,0}$) and $q_f (\equiv M_c/M_{2,0}$) are the mass ratios of the primordial and present day binaries, respectively. At the onset of (unstable) mass transfer, the ratio, $R_{1,0}/a_0$, is approximately equal to the Roche lobe radius of the progenitor primary divided by the primordial separation; its value is constrained by Roche geometry and can be expressed as a simple function of $q_0$ (see, Eggleton 1983). Given that the radius of the red giant will have a very strong dependence on $M_ c$ and a very weak dependence on $M_{1,0}$, the constraint of Roche geometry allows the approximation of $P_0$ by a function of $M_c$ (see, e.g., Rappaport et al. 1995).

Assuming a specific value of $\eta$, and using the observed orbital period of the binary ($P_f$) and the inferred mass of the present-day companion $M_1$ ($\equiv M_{2,f} = M_{2,0}$), strong limits can be imposed on the possible values of the white dwarf mass (we take $M_{\rm wd} = M_c +0.01 M_\odot$), by solving Eqn.~(\ref{eqn:CE2}) implicitly for $M_{\rm wd}$ and applying the constraint that $M_{1,0} > M_{2,0}$ (i.e., $q_0>1$). It is reasonable to suppose that $\eta$ will lie in the range 0.5 to 1.0. Taking the upper limit of this range for $\eta$, we find that the only acceptable values of the core mass for KIC 9164561 are $M_c \simeq 0.36- 0.38 \, M_\odot$ while for KIC 10727668 they are $M_c \simeq 0.40 - 0.42 \, M_\odot$. Assuming the lower limit of this range of $\eta$, we find that the only acceptable values of the core mass for KIC 9164561 are $M_c \simeq 0.39 - 0.40 \, M_\odot$, while for KIC 10727668 they are $M_c \simeq 0.44 - 0.45 \, M_\odot$. These ranges of values\footnote{The lower limits arise from the requirement that $q_0 >1$, by definition, while the upper limits correspond to $q_0 \lesssim 1.2$ to take into account the fact that we are considering only primaries of masses $\lesssim 2.5\,M_\odot$ so that only He WDs are produced.} of $M_c$ are clearly incompatible with the estimated values of $M_{\rm wd}$ for both KIC 9164561 and KIC 10727668 based on the cooling evolution of the white dwarfs (see the next section), as well as with the empirically determined value for KIC 9164561 from RVs.  We conclude that for a reasonable range of common envelope efficiencies, the formation of either or both of these systems via the CE channel is rather unlikely.

{\em (ii) The Algol Channel}

The other channel involves stable mass transfer during the evolution of the progenitor binary. In this scenario, the star that is more massive (the donor) evolves more quickly and fills its Roche lobe on its nuclear timescale before its less massive companion (the accretor) can do so. Even if the mass ratio is as large as $q_0 \simeq 2$, it is entirely possible for the mass transfer to proceed stably\footnote{Mass transfer could then proceed on a thermal timescale.}, if the donor has a substantial radiative envelope (Nelson \& Goliasch 2014). As an example of evolution that could explain the formation of the two observed systems, consider a 1.5 $M_\odot$ donor that has evolved beyond the TAMS and lies in the Hertzsprung gap when it fills its Roche lobe. This star can stably transfer mass to a 1.2 $M_\odot$ companion on approximately a nuclear and/or thermal timescale (Case AB evolution). At this mass transfer rate (of $\sim$$10^{-8} \,M_\odot$ yr$^{-1}$), the accretor is not likely to expand and fill its own Roche lobe\footnote{If it were to fill its own Roche lobe then this type of evolution would probably lead to either a contact binary or a merger.}.  During the mass transfer process, the donor can evolve up the red giant branch and form a substantial degenerate helium core. Since the radius of the giant has a very strong dependence on $M_c$ but a very weak dependence on its total mass, the giant will continuously expand with a concomitant increase in the core mass until the envelope surrounding the core is so depleted (to a mass of the order of 0.01 $M_\odot$; the remainder having been transferred to the companion star) that it cannot be sustained and ultimately collapses onto the remnant degenerate helium core. If approximately 1\,$M_\odot$ is transferred conservatively to the accretor during this process, the resulting system may then resemble either of the observed ones. Details concerning the possible progenitor evolutions will be discussed in a future paper (Nangreaves et al. 2015).

\subsection{Cooling Evolution}
\label{sec:cool}

Assuming that stable mass transfer from a red giant led to the formation of the bloated helium WDs in both KIC 9164561 and KIC 10727668, it is possible to check whether the observed properties of the WDs are consistent with cooling models of degenerate helium dwarfs. Moreover, if the properties of either WD are found to be consistent, it is then possible to use it as a chronometer in order to determine the age of the binary after the termination of mass transfer, i.e., just after the collapse of the giant to form a helium WD (HeWD).

The study of the formation and cooling of HeWDs from progenitor red giants has been carried out by a number of groups (see, e.g., Podsiadlowski et al. 2003; Nelson et al.~2004; Althaus et al.~2005, 2013; Istrate et al.~2014). We use the code described by Nelson et al.~(2004) to analyze the cooling evolution of very low-mass helium WDs.  It is assumed that the giant (i.e., the WD progenitor) had a solar metallicity ($Z=0.02$), and that its initial mass was between 1.0 and 1.5 $M_\odot$. The initial hydrogen abundance (by mass) of the progenitor star while on the ZAMS is taken to be equal to 0.71 and the ratio of the mixing length to the pressure scale-height ratio ($\ell/H_p$) is set equal to 1.5. The initial orbital separation was carefully selected so that only very low mass helium WDs were created. Diffusion (gravitational settling and radiative levitation) was not included in the computations because diffusion is a very fragile physical process that can be negated by turbulence in the outer layers of these rapidly rotating (proto-)white dwarf stars. Moreover, the inclusion of diffusion has little effect on the cooling times for the models of interest (i.e., bloated helium WDs) because these systems are very young and have not even evolved to reach their maximum surface temperatures before joining the usual WD cooling track in the HR diagram. It is only after the WDs have passed through their maximum $T_{\rm eff}$ that the compression of the outer hydrogen-rich layers, due to contraction, can cause the residual hydrogen to start to burn, leading to a rapid expansion of the outer layers as a result of a thermonuclear runaway (TNR).

\begin{figure}
\begin{center}
\includegraphics[width=0.49 \textwidth]{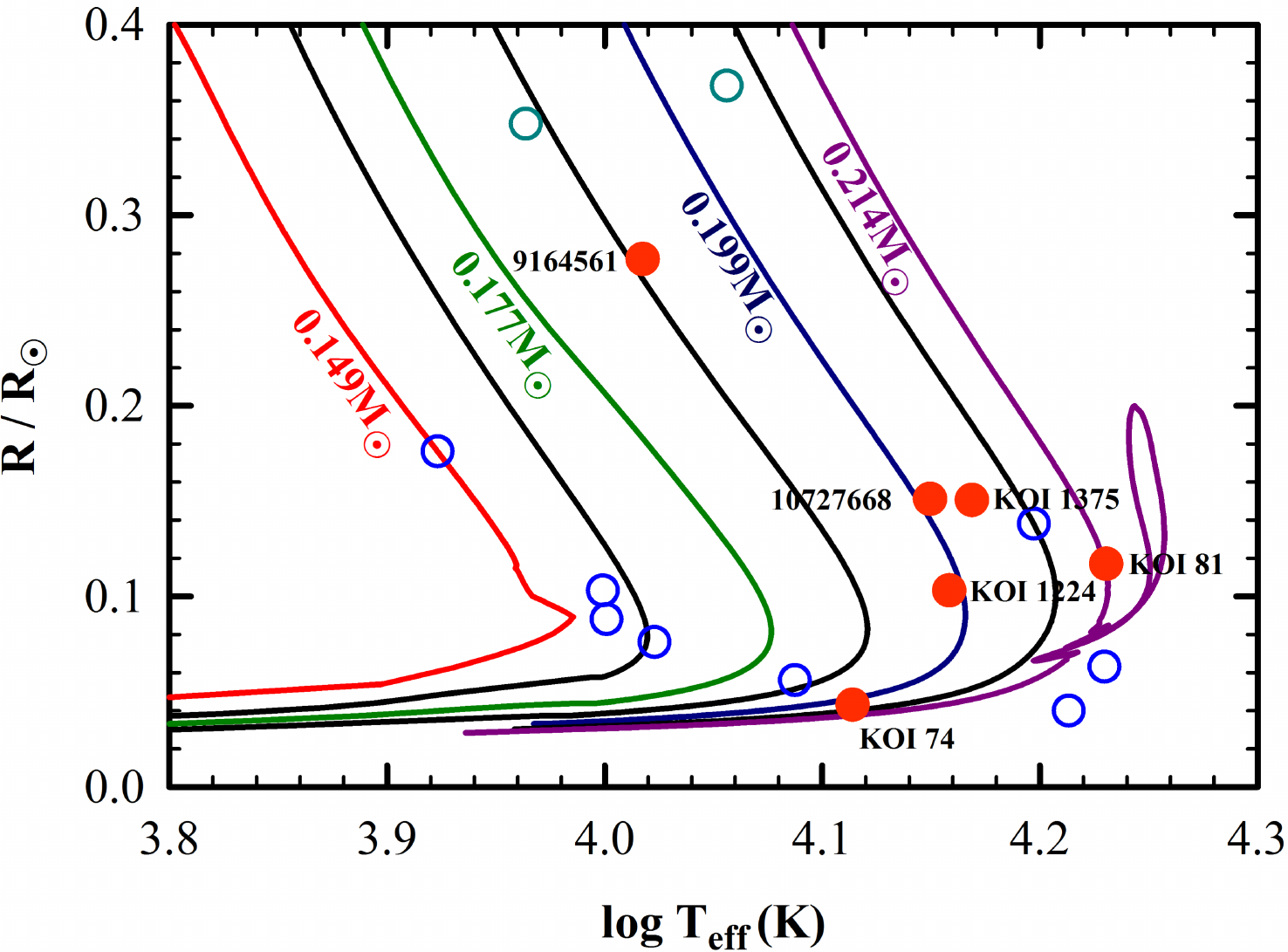} 
\caption{Cooling tracks in the radius--$T_{\rm eff}$ plane for low-mass white dwarfs that have lost their outer envelopes via mass transfer to a companion star.  The tracks are labeled by the mass of the white dwarf.  The filled red circles are the approximate current locations of the six known thermally bloated hot white dwarfs found among the {\em Kepler} binaries.  The open blue circles are eight low-mass white dwarfs found in double-degenerate systems (Hermes et al.~2014), while the open green circles are for systems similar to those reported here, and that were found using the WASP photometric database (Maxted et al.~2013; 2014); these have been added for comparison.}
\label{fig:cooling}
\end{center}
\end{figure}

In Figure \ref{fig:cooling}, we show the results of the cooling evolution of several models that have masses near or smaller than the TNR lower mass limit, i.e., $M_c \lesssim 0.21\,M_\odot$. The evolution of the radius of the WD is shown as a function of its effective temperature, $T_{\rm eff}$. When the envelope of the progenitor giant collapses, the radius of the proto-WD continually decreases, as long as no TNRs occur, until it asymptotically reaches its zero-temperature radius. The surface temperature of the object initially increases (this increasing temperature defines the proto-WD phase) until it reaches a maximum value of $T_{\rm eff}$ where it then follows a cooling track and asymptotically approaches zero temperature. 

Assuming that the white dwarfs in both KIC 9164561 and KIC 10727668 formed according to the scenario discussed above, we can infer reasonably precise values of the mass based on the observed radii and $T_{\rm eff}$ values superposed on the cooling tracks (see Fig.~\ref{fig:cooling}). We find that the mass of the WD in KIC 9164561 is $0.189 \pm 0.010\,M_\odot$, and that of the WD in KIC 10727668 is $0.201 \pm 0.010\, M_\odot$.  These masses agree with those inferred from the cooling evolutions of Althaus et al.~(2013) to within 5 to 10\% (their values of the mass being smaller), and to within $<2$\% for the values inferred from the work of Istrate et al.~(2014).  We also carried out this same type of WD cooling calculations with the MESA code (Paxton et al.~2011) and found very similar results.  The ages of these objects as measured from the cessation of mass transfer are approximately 925 Myr and 950 Myr, respectively. If the companion star in KIC 9164561 has a mass of $2.0 \, M_\odot$, then the time required to reduce its average density to the inferred value of 0.17 g cm$^{-3}$ is in good agreement with the theoretical predictions for the evolution of a solar metallicity star. The orbital period predicted by our models is approximately 1.7 days, but it should be noted that our assumptions about the degree of mass conservation during the mass-transfer phase and the magnitude of angular momentum losses leading to the formation of the giant are not precisely known, and this uncertainty could have a significant impact on the final value of the orbital period. As for KIC 10727668, the predicted orbital period from our models almost matches the observed one. However, the predicted age and somewhat higher mass of the companion star are very problematic because the observationally inferred average density implies that the companion star has not significantly evolved  from its initial position on the ZAMS. Based on the implied cooling age of the system and a mass of 2.2 $M_\odot$, we would expect the star to be much closer to the TAMS with a correspondingly lower density (contrary to what is implied in Figure 3). If the mass of the companion were substantially less (say, $\approx$1.7 $M_\odot$), then a WD composed primarily of helium and with a mass of $0.21\, M_\odot$ would be entirely compatible with the observations.

Based on the cooling evolution, we conclude that the theoretical models for the evolution of KIC 9164561 lead to properties that agree well with the observationally inferred properties. The predictions for KIC 10727668 are problematic, especially given the implied cooling age of the white dwarf in the system. For an age in the range of 100--200 Myr, a white dwarf mass of $0.27 \,M_\odot$ (inferred from the Doppler boosting effect; see Sect.~\ref{sec:DB}), and $T_{\rm eff} \simeq 14,000 \,K$, it is very difficult  to model the properties with any type of solar metallicity HeWD. To understand why this is true, consider the evolutionarily limiting case for which all of the hydrogen has been stripped from the atmosphere and envelope of the HeWD. Based on the WD cooling tracks of Nelson \& Davis (2000), a 0.27 $M_\odot$ WD of this type can reach a temperature of 14,000K at either of two ages: (i) considerably less than 10 Myr; or, (ii) $\approx 250$ Myr. The luminosity of the WD for the latter case is two orders of magnitude smaller than what is inferred (see Table \ref{tab:sys}) and can be automatically ruled out. For the other solution (corresponding to the early-WD phase before the maximum $T_{\rm eff}$ is reached), the luminosity is approximately correct but a cooling age of only several million years after the birth of the WD seems to be unacceptably small based on probability arguments. If a thick layer of hydrogen were to be added to the envelope, the cooling would be slowed but incompatibilities with the observations would persist. For example, the luminosity of the HeWD while it is still young would be considerably higher and it would likely undergo a TNR. Perhaps we are observing the HeWD while it is undergoing one of its TNR cycles; the probability of observing it at this epoch may be unacceptably small. Another possible explanation might be a difference in the actual metallicity from that which we assumed; however, making the metallicity smaller generally makes the incompatibilities larger. Thus if the mass of the WD in KIC 10727668 is actually about 0.27 $M_\odot$, then it might have been derived from a higher mass progenitor ($> 2.5 \, M_\odot$) or perhaps via a different evolutionary pathway than we have considered here. We plan to investigate all of these possibilities in a future paper.

For the sake of completeness, we have included the robust measurements of effective temperatures and radii of eight other eclipsing or tidally distorted, extremely-low mass WDs (found in double degenerate systems; see Hermes et al.~2014). The observations are plotted in Fig.~\ref{fig:cooling} as open blue circles. It should be noted that these additional WDs have properties that are reasonably similar to the six hot WDs found by Kepler except that they are typically less bloated. Based on the cooling models, we expect the masses of all of these systems to reside in the range of 0.15-0.25 $M_\odot$, which is generally consistent with the values reported in Hermes et al.~(2014).  Finally, in this regard, we note that these white dwarfs have a very different evolutionary scenario from the systems examined in our work in that they are likely the result of two common-envelope episodes. 

\subsection{Future Evolution}
\label{sec:future}
Here we examine very briefly what the future holds for systems such as KOI 74, KOI 81, KOI 1224, KOI 1375, KIC 9164561, and KIC 10727668.  When these systems evolve to the point where the primary A star fills its Roche lobe and commences mass transfer onto the white dwarf, their He core masses will range from 0.19 to 0.27$\, M_\odot$ for the shortest (KIC 9164561) and longest (KOI 81) orbital periods of these systems, respectively (as estimated using Eqn.~(\ref{eqn:Mwd_P2})).  Once the mass transfer onto the white dwarf begins, it will be dynamically unstable due to the very large mass ratio (typically 10:1), and a common envelope will ensue.  Using Eqn.~(\ref{eqn:CE}), we can compute an approximate expression for the ratio of the final orbital separation to the initial orbital separation, and we find: 
\begin{equation}
\frac{a_f}{a_i} \simeq \frac{M_{\rm wd} M_c}{M_e \,M_A}\, \frac{\eta \,r_L(q)}{2} \approx 0.003
\label{eqn:aoai}
\end{equation}
where $M_{\rm wd}$ is the mass of the hot white dwarfs being studied in these systems, $M_e$ and $M_c$ will be the envelope and core mass of the A star when it evolves to fill its Roche lobe, $M_A \equiv M_c+M_e$, $\eta$ is the common envelope parameter of order unity discussed above, and $r_L$ is the Roche-lobe radius of the A star in units of the orbital separation (with an expected numerical value of $\approx 0.6$). 

	If we apply Eqn.~(\ref{eqn:aoai}) to estimate the final orbital separation that will result from this upcoming common envelope phase, we find that  $a_f \lesssim 0.003 \,a_i$.  The current orbital separations of the six bloated hot white dwarfs in the {\em Kepler} field range from 6.4 to 48$\,R_\odot$.  From this we can infer that the final orbital separations will be $a_f \lesssim 0.02-0.14 \, R_\odot$, for KIC 9164561 to KOI 81, respectively.  
	
	All of these will be binaries comprised of two roughly equal mass white dwarfs each of $\approx 0.2\, M_\odot$, if the core of the A star does not merge with the initial low-mass white dwarf in the system.  The corresponding orbital periods and Roche-lobe radii would range from 3/4 to 14 min, and $0.007-0.05 \,R_\odot$, respectively.  All but the largest of these Roche-lobe radii are too small to harbor the white dwarf cores that will be unveiled by the common envelope.  
Thus, we conclude that the orbital periods of the group of double degenerate systems recently studied by Brown et al.~(2014) and Hermes et al.~(2014) result from much wider systems.  This is emphasized by the fact that the masses of the primary in the double degenerate binaries in the latter studies are typically much more massive than would result from the six binaries containing bloated hot white dwarfs in the {\em Kepler} field.

\section{Summary and Conclusions}

In this study we report two new low-mass, thermally bloated, hot white dwarfs in binary systems with companion A stars.  The systems are similar in their basic nature to those in KOI 74, KOI 81, KOI 1224, and KOI 1375.  This brings to six the total number of such systems in the {\em Kepler} field. Of the six systems, the two new systems have the shortest orbital periods at 1.27 d and 2.3 d, while KIC 9164561 has the largest, i.e., the most thermally bloated, of the six white dwarfs.  Its white dwarf has a radius that is 14 times larger than its fully degenerate radius---it is more than 1/2 $R_\odot$ in {\em diameter}!

The mass of the white dwarf in KIC 9164561 is 0.197 $\pm$ 0.005 $M_\odot$ and is determined from direct RV measurements of the primary A star as well as some inferences about the mass of the A star (i.e., the mean density and $T_{\rm eff}$ lead to an estimate for $M_1$).  By contrast, the mass of the white dwarf in KIC 10727668  is more uncertain at 0.27 $\pm$ 0.03 $M_\odot$.  Since there are no RV measurements of this fainter system, we relied on measurements of the Doppler boosting effect to determine the orbital velocity of the A star.  The accuracy of this latter mass is not quite sufficient for detailed comparisons with binary evolution or white-dwarf cooling models.  We strongly advocate that a larger telescope be used to determine an accurate RV for this system.

We have shown that KIC 9164561 is very plausibly part of a triple-star system with an outer period of $\gtrsim 8$ years.  The mass of the third star is likely constrained to be $\lesssim 0.8 \,M_\odot$.  

Binary evolution scenarios involving two low-mass, e.g., $\approx$ 1.1 and 1.5 $M_\odot$, primordial stars that undergo stable, conservative mass transfer can nicely explain the current mass and evolutionary state of both the companion star and the hot white dwarf in KIC 9164561.  For this system, we estimate an age of $\sim$925 Myr since the envelope of the white dwarf progenitor was transferred to the current A star.  By contrast, it is more difficult to explain the current state of both the A star and its white dwarf companion in KIC 10727668.  If the white dwarf actually has a mass as high as is indicated by the Doppler boosting technique (i.e., 0.27 $M_\odot$) then it is difficult to understand how the white dwarf can still be so thermally bloated.  If, on the other hand, we take the uncertainty seriously, and allow for a mass as low as, say, 0.22 $M_\odot$, then there is still a discrepancy between the inferred cooling age of $\approx$ 950 Myr, and the apparent location of the A star in the HR diagram which is still near the ZAMS.  

The growing statistical ensemble of these systems should make it worthwhile to (1) make renewed attempts to measure the RVs in four of the remaining bloated hot white dwarf systems in order to better determine $M_{\rm wd}$, and (2) directly model the binary evolution of such systems with a code that can evolve two stars {\em simultaneously}, e.g., MESA (Paxton et al.~2011).  

 \vspace{0.2cm}
 
 {\it Note Added In Manuscript}:  After this work was completed we became aware of an impressive collection of 18 similar bloated hot white dwarf systems found in the WASP photometric database (Maxted et al.~2013; 2014; and references therein).  Two of these have accurate measurements of the mass, radius, and $T_{\rm eff}$ for the bloated hot white dwarf and show multi-periodic pulsations. These are now also shown in Fig.~\ref{fig:cooling}.

\line(1,0){200}

\acknowledgements We are grateful to the entire {\it Kepler} team for making this study possible.  We thank Jon Swift for taking exploratory spectra of KIC 10727668 to see if RV measurements can be acquired in the future, and Dave Latham for his early contribution to this project.  The RV observations were made with the Nordic Optical Telescope, operated by the Nordic Optical Telescope Scientific Association at the Observatorio del Roque de los Muchachos, La Palma, Spain, of the Instituto de Astrofisica de Canarias.  We gratefully acknowledge that two of the RV points were acquired by Pilar Monta\~nes-Rodr\'iguez, and an additional one was obtained during time allocated to Hans J. Deeg.  We thank Antonio Frasca of Catania Astrophysical Observatory for some important tips in combining the model spectra.  Mattia Gandolfi was quite inspirational for this work.  We are grateful for the CAT and NOT service time awarded to this project, and to the staff members at the NOT for their valuable and unique support during the observations.  This research was supported in part by the MINECO ESP2013-48391-C4-2-R grant.  L.N. thanks the Natural Sciences and Engineering Research Council (NSERC) of Canada for financial support, Calcul Qu{\'e}bec for providing computing facilities, and Ernie Dubeau and Heather Nangreaves for technical assistance.  R.S.O.\ acknowledges NASA support through the Kepler Participating Scientist Program.  This research has made use of the NASA Exoplanet Archive, which is operated by the California Institute of Technology, under contract with the National Aeronautics and Space Administration under the Exoplanet Exploration Program. The data presented in this article were obtained from the Mikulski Archive for Space Telescopes (MAST). STScI is operated by the Association of Universities for Research in Astronomy, Inc., under NASA contract NAS5-26555. Support for MAST for non-HST data is provided by the NASA Office of Space Science via grant NNX09AF08G and by other grants and contracts.  We made use of J-band images that were obtained with the United Kingdom Infrared Telescope (UKIRT) which is operated by the Joint Astronomy Centre on behalf of the Science and Technology Facilities Council of the U.K.

\end{document}